\documentclass[11pt,a4paper]{article}
\pdfoutput=1
\usepackage[pdftex]{graphics}
\usepackage{jheppub}
\usepackage{amsmath,amssymb,amsfonts}
\usepackage{enumitem}
\usepackage{multirow}
\usepackage{array,booktabs}
\usepackage{tikz}
\usepackage{slashed}
\usepackage{verbatim}
\usepackage{float}
\usepackage{graphicx}
\usepackage{subcaption}
\usepackage{blkarray}
\usepackage{empheq}
\usepackage{cancel}

\newcommand{\eq}{\begin{equation}}
\newcommand{\eqe}{\end{equation}}
\newcommand{\eqa}{\begin{eqnarray}}
\newcommand{\eqae}{\end{eqnarray}}
\newcommand{\nn}{\nonumber}
\newcommand{\bn}{\begin{enumerate}}
\newcommand{\en}{\end{enumerate}}
\newcommand{\bl}{\begin{align}}
\newcommand{\el}{\end{align}}
\def\ie{\begin{equation}\begin{aligned}}
\def\fe{\end{aligned}\end{equation}}

\def\a{\alpha}

\def\u{\upsilon}

\def\det{{\rm det}}

\def\jmath{{j}}
\def\bl#1\el{\begin{align} #1 \end{align}}
\def\bg#1\eg{\begin{gather} #1 \end{gather}}

\def\bld#1\eld{\begin{aligned} #1 \end{aligned}}
\def\bgd#1\egd{\begin{gathered} #1 \end{gathered}}

\def\bE{{\mathbb{E}}}

\usetikzlibrary{snakes}
\usetikzlibrary{shapes.misc}
\usetikzlibrary{decorations.markings}

\tikzset{line/.style={line width=0.25mm},
curve/.style={line,smooth,tension=1},
->-/.style={decoration={
  markings,
  mark=at position #1 with {\arrow[>=stealth]{>}}},postaction={decorate}},
-<-/.style={decoration={
  markings,
  mark=at position #1 with {\arrow[>=stealth]{<}}},postaction={decorate}},
}

\newcommand*\widefbox[1]{\fbox{\hspace{1em}#1\hspace{1em}}}

\title{Into the EFThedron and UV constraints from IR consistency}
\author[1]{Li-Yuan Chiang}
\author[1,2]{Yu-tin Huang}
\author[1]{Wei Li}
\author[1]{Laurentiu Rodina}
\author[1]{He-Chen Weng} 

\affiliation[1]{Department of Physics and Center for Theoretical Physics, National Taiwan University, Taipei 10617, Taiwan}
\affiliation[2]{Physics Division, National Center for Theoretical Sciences, Taipei 10617, Taiwan}
 
\emailAdd{yutinyt@gmail.com}

\abstract{Recently it was proposed that the theory space of effective field theories with consistent UV completions can be described as a positive geometry, termed the EFThedron. In this paper we demonstrate that at the core, the geometry is given by the convex hull of the product of two moment curves. This makes contact with the well studied bi-variate moment problem, which in various instances has known solutions, generalizing the Hankel matrices of couplings into moment matrices. We extend these solutions to hold for more general bi-variate problem, and are thus able to obtain analytic expressions for bounds, which closely match (and in some cases exactly match) numerical results from semi-definite programing methods. Furthermore, we demonstrate that crossing symmetry in the IR imposes non-trivial constraints on the UV spectrum. In particular, permutation invariance for identical scalar scattering requires that any UV completion beyond the scalar sector must contain arbitrarily high spins, including at least all even spins $\ell\le28$, with the ratio of spinning spectral functions bounded from  above, exhibiting large spin suppression. The spinning spectrum must also include at least one state satisfying a bound $m^2_{J}<M_h^2  \frac{( J^2{-}12) ( J^4 {-} 32 J^2 {+}204)}{8 (150{-}43 J^2 {+}2 J^4)}$, where $J^2=\ell(\ell{+}1)$, and $M_h$ is the mass of the heaviest spin 2 state in the spectrum.}

\begin{document}
\maketitle

\newpage
\section{Introduction}
Effective field theories (EFT) are a convenient tool to parameterize our ignorance of the UV physics and at the same time provide a useful description of the IR. Precisely because of its role of connecting the UV with the IR, it has long been a fertile playground for understanding how consistency conditions on one side impose non-trivial constraints on the other. A prominent example is the positivity of the leading four derivative couplings in EFT~\cite{Adams:2006sv} relating to the $a$-theorem~\cite{Komargodski:2011vj} as well as the Weak Gravity Conjecture~\cite{ArkaniHamed:2006dz}.   

In the original work~\cite{Adams:2006sv}, positivity bounds were derived from the forward limit of the low energy amplitude, which is related to the cross-section of the UV scattering amplitude through dispersion relations and the optical theorem. As one extends away from the forward limit, the fact that the imaginary contribution in the dispersion relation has a positive partial wave expansion further constrains the structure of the derivative expansion. This was first proposed in~\cite{Nicolis:2009qm}, and explored in a more systematic framework recently~\cite{Bellazzini:2017fep, deRham:2017avq, Bellazzini:2020cot, Arkani-Hamed:2020gyp}. For identical states, crossing symmetry drastically constrains the allowed space of EFT couplings, in general reducing it to finite compact regions\cite{Tolley:2020gtv, Caron-Huot:2020cmc, Caron-Huot:2021rmr, Sinha:2020win}. Monodromy relations push this reduction even further, seemingly converging to a single point corresponding to the low energy expansion of the open superstring \cite{Huang:2020nqy}.

A geometric characterization of the space of consistent EFTs was recently given in~\cite{Arkani-Hamed:2020blm}, where positivity bounds are given as the boundaries of the EFThedron, which is a product of convex spaces. To be concrete, let us begin with the low energy scattering amplitude, which can be parametrized as
\eq
M(s,t)^{\text{IR}}=\sum_{k,q} g_{k,q} s^{k-q} t^q\,,
\eqe
where the Taylor expansion in $s$ and $t$ reflects the derivative expansion for the low energy description, and the couplings $g_{k,q}$ are the Wilson coefficients defined on an on-shell basis. Via dispersion relations, the couplings admit a representation as a positive sum over data provided by the UV spectrum. In particular
\eq\label{Xeq}
g_{k,q}=\sum_i p_i \frac{1}{m_i^{2k+2}} X_{\ell_i,k,q}\,, 
\eqe
where $i$ labels the UV state, and $\{p_i,m_i, \ell_i\}$ are the spectral function, mass and spin of state $i$. The coefficient $X_{\ell_i,k,q}$ is related to the Taylor expansion of the Legendre polynomial $P_{\ell}(x)$, as well as the precise channel ($s$ or $u$) in which the state was exchanged. 

Importantly, as unitarity requires the spectral function $p_i$ to be positive, the dispersive representation of the couplings is in fact giving a convex hull. In~\cite{Arkani-Hamed:2020blm} the geometry of the hull was explored in full for the $s$-EFThedron, corresponding to the scenario where the UV states appear only in the $s$-channel. There are no $t$-channel poles since the dispersion relation is implemented for fixed $t$.  The geometry was found to be the product of a bounded moment curve and a cyclic polytope, whose individual boundaries are well known. For the product, the new boundaries involve the total positivity of the following Hankel matrix
\eq
H(g)=\begin{pmatrix}
g_0& g_1& g_2\\
g_1& g_2 & g_3&\ldots\\
g_2& g_3 & g_4\\
&\vdots&&\ddots
\end{pmatrix}\ge 0\,,
\eqe
where here $g_k$ represents the inner product of $\vec{g}_{k}=(g_{k,0},g_{k,1},\cdots)$ with any boundary of the cyclic polytope. However, as we will see, these are in fact incomplete.  For the $su$-EFThedron, where both $s$- and $u$-channel thresholds are present, the cyclic polytope is replaced by a more complicated object, and only a partial exploration was done. One of the goals of this paper is to fill the gaps for both geometries, by showing they both originate from a common structure, the product moment curve.

In this paper, while we focus on the case of four-dimensions, using Legendre polynomials for external scalars and Wigner $d$-matrix for spinning states, the results are applicable to higher dimensions in two senses. One, the transformation of the dispersion relation to a double moment problem can be straightforwardly carried out in higher dimensions by merely substituting Legendre with Gegenbauer polynomials. Second, the four-dimensional bounds can be viewed as a weaker set of constraint for higher dimensions, since we can always put the external states on a four-dimensional sub-plane, and the higher dimensional basis must be positively expanded on the four-dimensional counterpart. 

When strictly in four-dimensions, complications arise from infrared divergences. In such a case, our bounds apply to effective theories of UV completions for which the cubic massless interactions are weakly interacting, and thus IR effects are higher order in couplings. Massless loops from the higher dimension operators will introduce logarithmic running, whose effect can be incorporated by modifying the definition of EFT couplings~\cite{Bellazzini:2020cot, Arkani-Hamed:2020blm}.

\begin{figure}[h]
\centering
 \includegraphics[scale=0.26]{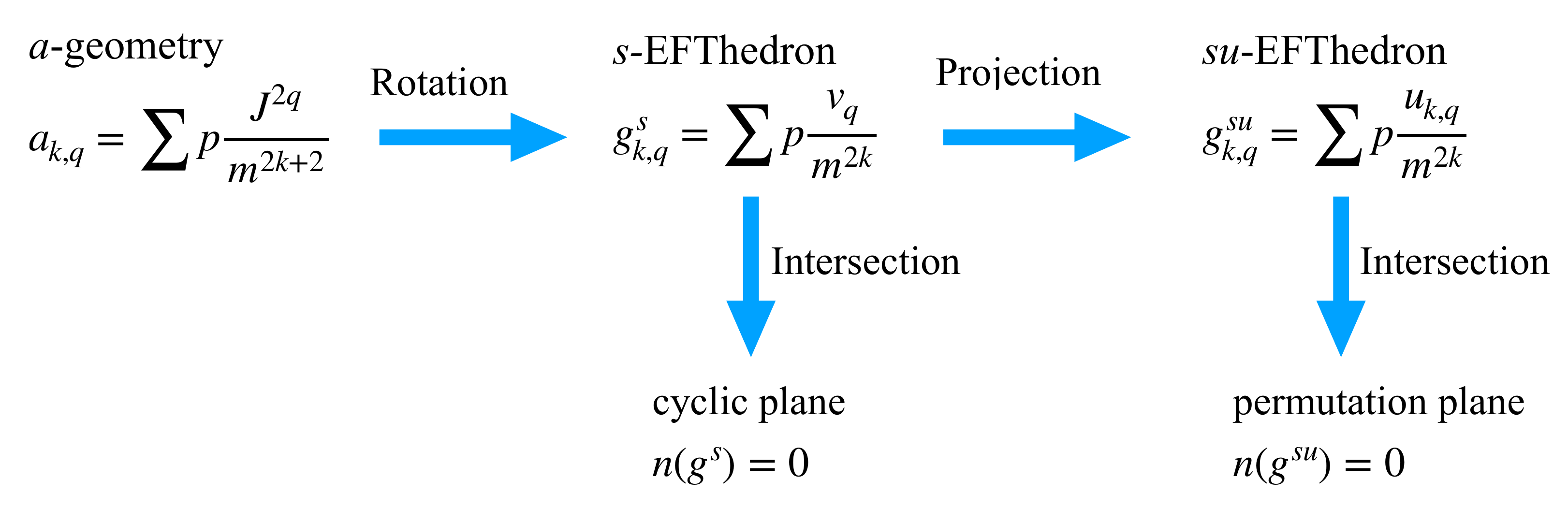} 
 \caption{Deriving $s$- and $su$-EFThedrons from $a$-geometry}
 \label{pic1}
\end{figure}

\paragraph{EFThedron from the moment problem}
We begin by demonstrating that even the $s$-EFThedron can be expressed in simpler terms. By a GL transformation, the cyclic polytope defined by $v_{l,q}$, which are complete polynomials in $J^2=\ell(\ell+1)$ of degree $q$, can be rotated into a discrete moment curve in $J^2$. Therefore, instead of considering the $g$-geometry defined by eq.(\ref{Xeq}), we can instead consider what we call the $a$-geometry, \eqa\label{Main}
a_{k,q}=\sum_i p_i \frac{1}{m_i^{2k+2}} J_i^{2q}\,.
\eqae
and return to the $g$-geometry by a simple rotation of the couplings. For the full EFThedron, the geometry is given by the Minkowski sum of the hulls arising from the $s$- and $u$-channels. Because the $u$-channel contribution is just itself a GL rotation of the $s$-channel one, it turns out we can obtain the full EFThedron geometry by a similar rotation, followed by projecting out couplings $g_{k,q}$ with $k+q=\textrm{odd}$, reflecting the $s-u$ channel symmetry. Thus we see both EFThedrons are unified into the $a$-geometry, as illustrated in fig.\ref{pic1}.

The significance of this result is that the $a$-geometry is explicitly the product of two moment curves. The complete characterization of its boundaries can then be formulated as the  ``bi-variate moment problem", which is a rich and well studied problem in mathematics~(see ~\cite{Moment} for an overview). The problem lies in finding both necessary and sufficient conditions that determine whether a set of ``moments" (in our terms -- couplings) can be expressed as eq.(\ref{Main}). Solutions to various cases of the moment problem are well established in the infinite dimension limit, which for us corresponds to the limit where we have access to Wilson coefficients of all derivative orders, that is arbitrarily high orders in $k$ and $q$. This allows us to systematically find all the boundaries of the $a$-geometry in the infinite dimension limit. The crucial conditions that we find, compared to~\cite{Arkani-Hamed:2020blm}, involve the generalization of the Hankel matrix into a moment matrix
\eq \label{bmatrix}
M(a)=\begin{pmatrix}
a_{0,0}& a_{1,0}& a_{0,1}& a_{2,0}\\
 a_{1,0}& a_{2,0}& a_{1,1}& a_{3,0}\\
 a_{0,1}& a_{1,1}& a_{0,2}& a_{2,1}&\ldots\\
 a_{2,0}& a_{3,0}& a_{2,1}& a_{4,0}\\
 &&\vdots&&\ddots
\end{pmatrix}
\eqe
which we require to be positive semi-definite. Having both necessary and sufficient conditions implies the EFThedron is in fact equivalent to the dispersion relations encoding locality, unitarity and analyticity!

In practice, we will only be privy to a finite number of Wilson coefficients, meaning we must truncate at some order $k\le m$. In such case the geometry arising from eq.(\ref{Main}) is referred to as the ``truncated moment problem". A solution to this problem is known as the positive extension \cite{Curto,article}, which involves finding an extension to the finite moment (or Hankel) matrices, which maintains the positivity. For the single moment problem, only a single extension is required, i.e. to find the solution for a coupling space up to some order $k\le m$, we need to impose Hankel matrix positivities up to order $k=m+1$, and then project out the $k=m{+}1$ couplings. Our particular bivariate moment problem involving the tensoring of a bounded with discrete moment has however not been previously studied in the literature. Motivated by the solution for the full bi-variate moment, we propose the set of constraints for $k=m$ in eq.(\ref{finaleq}) for the $a$-geometry. Explicit analysis shows that by including positive extension to $k=m{+}1$, the region carved out by these constraints converges rapidly to the numerical result, while in some cases give the analytic exact boundary. Thus we conjecture that the constraints given in eq.(\ref{finaleq}), augmented with sufficient degree of positive extension yields the complete boundary.

In the second row in Figure \ref{pic1}, we considered further imposing the cyclic or permutation invariance of the low energy four point amplitude. These constraints imply linear relations between the couplings $g_{k,q}$ or $a_{k,q}$, which define what we will call null or symmetry planes. Importantly, once we require the EFThedron to intersect with the symmetry plane, the resulting space becomes bounded in all directions, as was also observed in~\cite{Caron-Huot:2020cmc, Sinha:2020win} by using the dispersion relation form of null constraints. The intersection allows us to project the typically high dimension geometry of the EFThedron to a low dimensional finite space. To clarify, we propose that cyclic or permutation invariance can be fully taken into account simply by slicing the EFThedron by symmetry planes, without having to assume any explicit dispersion relation form of the null constraints. All bounds are obtained by working exclusively at the level of couplings, that is with inequalities of the form $f(a_{k,q})\ge 0$, intersected with symmetry planes $n(a_{k,q})=0$. The equivalence is guaranteed by having the necessary and sufficient conditions for eq.(\ref{Main}) to have valid solutions.

\begin{figure}[h] 
 \centering
  \includegraphics[width=0.9\textwidth]{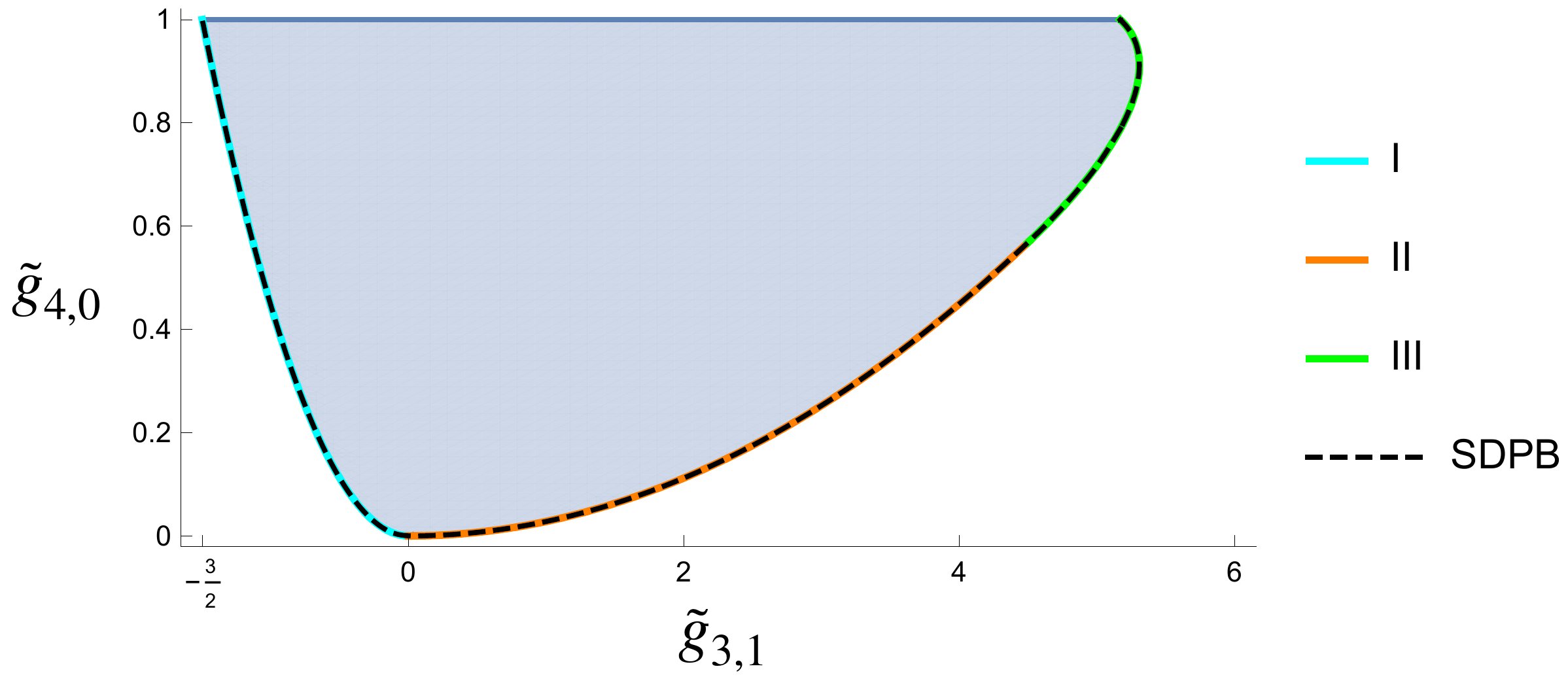} 
   \caption{($\tilde{g}_{3,1}$, $\tilde{g}_{4,0}$) space with the $k=4$ null constraint, geometry vs SDPB}
\label{introfig}
\end{figure}

We compare the boundaries with that obtained from numerical semidefinite programing using SDBP package~\cite{Simmons-Duffin:2015qma, Landry:2019qug}. In particular we compute the analytic structure of the allowed region relevant for $g_{3,1}$ and $g_{4,0}$ of the $s{-}u$ channel EFThedron, after intersecting with the first symmetry plane, which is at $k{=}4$. By imposing constraints involving up to $k=5$ couplings, we obtain the following result:
\begin{align}
\nonumber \textrm{Region I:}\quad \tilde{g}_{31} &=-\frac{3}{2} \sqrt{\tilde{g}_{40}},  &&0 \leq \tilde{g}_{40} \leq 1 \\ 
\nonumber\textrm{Region II:}\quad \tilde{g}_{31}&=\frac{1}{2} \sqrt{\frac{427}{3}  \tilde{g}_{40}}, &&0 \leq  \tilde{g}_{40} \leq \frac{243}{427}\\ 
\textrm{Region III:}\quad  \tilde{g}_{31}&=\frac{30}{7}  \tilde{g}_{40}+\frac{37}{42} \sqrt{ \tilde{g}_{40}\left(21-20  \tilde{g}_{40}\right)},  &&\frac{243}{427} \leq  \tilde{g}_{40} \leq 1
\end{align}
where we have defined $ \tilde{g}_{3,1}\equiv \frac{g_{3,1}M^2}{g_{2,0}}$ and $ \tilde{g}_{4,0}\equiv \frac{g_{4,0}M^4}{g_{2,0}}$, with $M$ the mass gap of the theory. We plot this in Figure \ref{introfig}, finding perfect agreement with numerical results from SDPB. In particular, we find an exact value for the bound
\eq
2\tilde{g}_{3,1}<\frac{9}{2} + \frac{7}{4} \sqrt{\frac{61}{5}}\approx 10.6125
\eqe 
identical to the value given in~\cite{Caron-Huot:2020cmc}. We've also demonstrated that these boundaries are tighter than previous proposal for s-channel EFT hedron~\cite{Arkani-Hamed:2020blm}.

\paragraph{UV constraints from IR symmetry}
The reason ``slices" of the EFThedron are finite is simple: the symmetry plane is always defined for fixed derivative order ($k$), and only the ``$J^2$" moment curve at different ($q$) orders is relevant for the intersection. The intersection geometry is therefore always between a polytope and a subplane, and one can easily see that a minority of vertices, defined by spin $\ell$, lies on one side of the plane. Thus the fact that the physical spectrum must have an empty image in the directions orthogonal to the symmetry plane, conventionally termed ``null constraints", requires a balancing act between the low and the higher spin part of the spectrum. 

Indeed we find null constraints, when now expressed in their dispersive representation form, lead to conditions on the UV spectrum. In both $s$- and full EFThedron, if the UV completion involves non-scalar states, cyclic and permutation invariance require the presence of an infinite number of massive spin states. Permutation invariance in fact requires every even spin $\ell\le 28$ to be present. While any massive UV completion will inevitably produce states of all spins via the discontinuity of the two-particle threshold, if we assume that the UV theory has a dimensionless coupling, for example  $g_s$ in string theory, such that massive loop corrections can be suppressed, this constraint becomes non-trivial. Furthermore, we also find upper bounds for the ratio of averaged spectral functions, defined as 
\eq
\langle p_{k,\ell}\rangle\equiv\sum_{\{i,\ell_i=\ell\}}\frac{p_{\ell_i}}{m^{2(k{+1})}_i}\,.
\eqe
For example, for identical scalars we find
\begin{align}
&\frac{\langle p_{4,\ell}\rangle}{\langle p_{4,2}\rangle}\leq\frac{12}{\ell (\ell+1) \left(\ell^2+\ell-8\right)},~(\ell\geq4)\notag\\
&\frac{\langle p_{5,\ell}\rangle}{\langle p_{5,2}\rangle}\leq\frac{216}{\ell (\ell+1) (\ell (\ell+1) (2\ell (\ell+1)-43)+150)},~(\ell\geq4)\,.
\end{align}
The result for distinct $k$ can be plotted as follows
$$\includegraphics[scale=0.5]{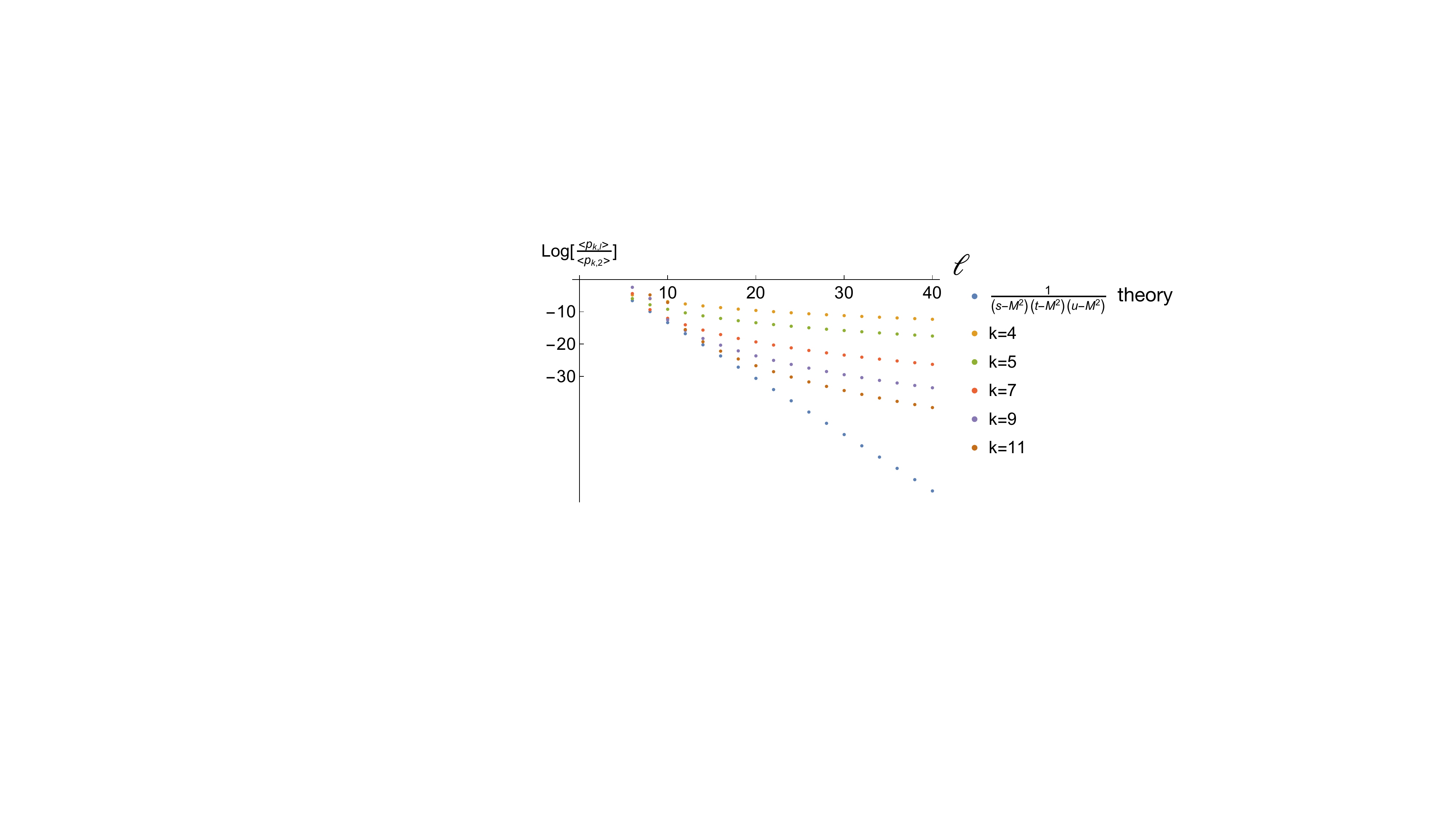}$$
Thus we see high spin suppression, at the level of finite spins, observed in~\cite{Arkani-Hamed:2020blm} and recently explored in~\cite{Bern:2021ppb}, can be in part attributed to the combination of permutation invariance and unitarity. 

Similarly, the mass distribution of the higher spin spectrum is constrained. Taking the heaviest spin-$2$ state to have mass $M_h$, then we find amongst the higher spin states, there must exists one satisfying
\eq
m^2_\ell \le M_h^2\frac{( J^2-12) ( J^4  - 32 J^2 +204)}{8 (150 - 43 J^2 + 2 J^4)}\,.
\eqe
The above results are just the simplest consequences of a small subset of null constraints. The full geometry of $\{p,\ell,m\}$ is highly non-trivial, as it must in fact satisfy an infinite number of null constraints. We expect many more conditions can be obtained by a systematic analysis of this geometry.

The paper is organized as follows. In Section \ref{review} we review the positive geometry of moment curves, and present our proposed complete solution to the bi-variate moment problem. In Section \ref{sec3} we show the $s$-EFThedron can be obtained by a GL rotation from the bi-variate moment form. We then present results for the geometry, which we compare to a functional approach. In Section \ref{sec4} we perform a similar analysis for the $su$-EFThedron. Besides a similar rotation, one also further needs to project out some couplings in order to satisfy $s{-}u$ channel symmetry. We similarly compare results to a functional approach. In Section \ref{UV} we begin an investigation into conditions imposed by null constraints on the UV spectrum itself. We derive constraints on spins, spectral functions, and masses. We conclude in Section \ref{conc}.

\section{Positive geometry of moment curves}\label{review}
In this section we give an overview of the geometry associated with the convex hull of moment curves. The convex hull is the space associated the positive sum of a fixed set of vectors $\{v_i\}$
\eq
\textrm{Conv}[\{v_i\}]\equiv \left\{ \sum_i^N p_i v_i \,\Big{|}\, p_i\geq0 \right\}\,.
\eqe
If a point is inside the hull, then there must be a corresponding solution $\{p_i\}$.  A moment curve is simply a curve whose image takes the form  $(1,x,x^2,\cdots)$. The convex hull of points on the moment curve is a space parameterized by the coordinates $\{a_0,a_1,\ldots\}$, satisfying
\eq\label{momentdef2}
a_k=\sum_i^N p_i x^k_i\rightarrow \begin{pmatrix}a_0\\a_1\\a_2\\\vdots\end{pmatrix}=\sum_i p_i \begin{pmatrix}1\\x_i\\x_i^2\\\vdots\end{pmatrix}\,,
\eqe
Stated in integral form, 
\eq\label{momentdef}
a_k=\int x^k d\mu
\eqe
where $d\mu$ represents a positive measure. For the purposes of this paper, the discrete and integral representations are equivalent. 

We are therefore interested in finding the necessary and sufficient conditions on $a_k$ such that they satisfy eq.(\ref{momentdef2}). Said in another way, what are the constraints that ensure the existence of a solution $\{p_i, x_i\}$ that solves eq.(\ref{momentdef2})? For a familiar application of this problem,  $a_1=\sum_i^n p_i x_i$ can be interpreted as the coordinate for the center of mass, for a system of $n$ particles of masses $p_i$, placed at positions $x_i$. The possible location for the center of mass is given by a convex region defined by the ``outermost" particles. The geometry we are looking for is a slight generalization of this problem, and is known as the ``moment problem". This comes in different versions, depending on the support of the variable $x$. the usual cases considered are  $x\in \mathbb{R}$,\, $x\in  \mathbb{R}_+$, \,$x\in [0,1]$ and finally $x$ taking only discrete values in some region. A recent detailed overview of the moment problem can be found in \cite{Moment}.

Eq.(\ref{momentdef2}) contains just one variable, $x$, and is termed the single or uni-variate moment problem. In general, one can have any number of moments. The double or bi-variate moment problem in particular has been extensively studied, and will be the main interest of this paper. In this case the moments are organized as
\eq\label{Prod}
a_{k,q}=\sum_i^N p_ix_i^k y_i^q \quad \rightarrow \left(\begin{array}{cccc}a_{0,0} & a_{0,1} &  a_{0,2} &  \\ a_{1,0} & a_{1,1} &  a_{1,2} & \cdots  \\  a_{2,0} & a_{2,1} &  a_{2,2} & \\ & \vdots &  &  \ddots \end{array}\right) = \sum_i^N p_i \left(\begin{array}{c} 1 \\ x_i \\ x_i^2 \\ \vdots\end{array}\right)\otimes \left(\begin{array}{cccc}1 & y_i & y^2_i & \cdots\end{array}\right)\,,
\eqe
where as before $p_i\ge 0$, while $x_i,y_i$ can have any domain. The physical geometry that is our main focus will have $x_i\in [0,1]$ (related to the ratio of the masses of the UV states to the cutoff), $y_i=\ell_i(\ell_i{+}1)$, with $\ell_i\in \mathbb{N}$ (corresponding to the spins of the UV states). Finally, the complete geometry will be given by intersecting the convex hull with symmetry planes corresponding to cyclic or permutation invariance.

Typically, the moment problem is formulated in the infinite dimensional limit, that is when one is interested in the space of an infinite sequence $a=\{a_0,a_1,\ldots\}$. However, for obvious practical purposes, we would like to carve out a finite dimensional space,  $a=\{a_0,a_1,\ldots, a_n\}$, using only a finite number of constraints. This setup is known as the truncated moment problem, and is considerably more complicated than the infinite moment problem, which in fact follows as a corollary whenever a truncated solution is known. In the single moment problem, the truncated solution is well established through the positive or flat extension theorems  \cite{Curto}, which are conjectured to hold also in the bi-variate case \cite{truncateconj}.

To our knowledge, for our particular case of physical interest, that is the bi-variate moment with one discrete and one continuous variable, a solution has not been studied. In the following we present a systematic way to find conditions for this problem in the infinite dimension limit. We note that some of the conditions have already been described in \cite{Arkani-Hamed:2020blm}, but we find new ones as well, and we claim these conditions are necessary and sufficient in the infinite dimensional limit. In Sections \ref{sec3} and \ref{sec4} we study two truncated cases, and we find these conditions quickly converge to the numerical results found by linear programming methods.

\subsection{Conditions for the full moment problem}
We begin by asking what are the necessary and sufficient conditions an infinite sequence $a=\{a_0,a_1,\ldots\}$ must satisfy to be a moment sequence, that is
\eq
a_k=\sum_i p_i x_i^k, \quad p_i\ge 0, \quad x_i\in\mathbb{R}
\eqe
It is well known the solution to this problem is given in terms of Hankel matrices, that must be positive semidefinite (PSD):
\eq\label{start}
H=\begin{pmatrix}
a_0&a_1&a_2&\\
a_1&a_2&a_3&\cdots\\
a_2&a_3&a_4&\\
&\vdots&&\ddots
\end{pmatrix}\ge 0
\eqe

Further conditions must be imposed if there are further restrictions on $x$. For example, one can have $x\in \mathbb{R}_+$, $x\in[0,1]$, or $x\in S$, where $S$ is some (ordered) discrete set. We will find uses for all such cases. The extra conditions one needs to impose are in terms of so called ``localizing matrices". For instance, to impose $x\ge0$, besides eq.(\ref{start}), we need the ``shifted" Hankel matrix:
\eq
H^\textrm{shift}=\begin{pmatrix}
a_1&a_2&a_3&\\
a_2&a_3&a_4&\cdots\\
a_3&a_4&a_5&\\
&\vdots&&\ddots
\end{pmatrix}\ge 0
\eqe

Next, for $x\le 1$, we have the ``twisted" Hankel
\eq
H^\textrm{twist}=\begin{pmatrix}
a_0-a_1&a_1-a_2&a_2-a_3&\\
a_1-a_2&a_2-a_3&a_3-a_4&\cdots\\
a_2-a_3&a_3-a_4&a_4-a_5&\\
&\vdots&&\ddots
\end{pmatrix}\ge 0
\eqe

Finally, if $x\in \{s_i\}$, for ordered $s_i$, this is known as the cyclic polytope, and the well known conditions are in terms of polytopal walls with vertices $s_i,s_{i+1},s_j,s_{j+1},\ldots$. For $\textrm{dim}[\vec{a}]=2d+1$ we have
  \begin{equation}
        \langle \vec{a}, \vec{s}_{i}, \vec{s}_{i+1} ,\vec{s}_{j}, \vec{s}_{j+1},\ldots \rangle 
        =\underbrace{
        \left|
        \begin{array}{cccccc}
        a_0 & 1 & 1  & 1 & 1 &\\
        a_1 & s_{i} & s_{i+1} & s_{j} & s_{j+1}  & \\
        a_2 & s_{i}^2 & s_{i+1}^2  & s_{j}^2 & s_{j+1}^2& \cdots \\
        a_3 & s_{i}^3 & s_{i+1}^3  & s_{j}^3 & s_{j+1}^3& \\
  & & \vdots &  &  & \ddots \\
        \end{array}
        \right|}_{2d+1} \geq 0.
    \end{equation}
while for $\textrm{dim}[\vec{a}]=2d$ we have  
\eqa
 \nonumber    \langle\vec{0}, \vec{a}, \vec{s}_{i}, \vec{s}_{i+1}, \vec{s}_{j}, \vec{s}_{j+1}, \ldots  \rangle&\ge&0 ,\\
    \langle \vec{a}, \vec{s}_{i}, \vec{s}_{i+1},\vec{s}_{j}, \vec{s}_{j+1}, \ldots,\vec{\infty} \rangle&\ge&0 ,
\eqae
where $\vec{0}=(1,0,\ldots,0)$ and $\vec{\infty}=(0,0,\ldots,1)$.

In the next section we present an approach that allows to systematically derive such constraints. The approach then generalizes easily for the double moment problem.

\subsubsection{Single moment}
We have seen that the sufficient condition for $\vec{a}$ to reside in the hull is that the Hankel matrix is positive semidefinite. Observing that an infinite dimensional Hankel matrix can be written as
\begin{equation}\label{aab}
    H(a) = \sum_i p_i \mathbf{x}_i \mathbf{x}_i^T = \sum_{i} p_i 
    \left(\begin{array}{c}
        1 \\ x_i \\ x_i^2 \\ \vdots
    \end{array}\right)
    \left(\begin{array}{cccc}
    1 & x_i & x^2_i & \cdots
    \end{array}
    \right),
\end{equation} 
we immediately see that the Hankel matrix is positive semi-definite, since for any vector~$\mathbf{v}$,
\begin{equation}
    \mathbf{v}^T H \mathbf{v} = \sum_i p_i (\mathbf{v}^T \mathbf{x}_i)^2 \geq 0\,.
\end{equation}
This implies all the principle minors are non-negative. 

Note that at this point, the Hankel matrix being positive-semidefinite is a necessary condition, but not yet sufficient. To show that it is sufficient we must show that satisfying the Hankel constraints implies the existence of a positive solution \(\{p_i\}\) that gives us \(\{a_k\}\). First, notice that \(f(x) = \mathbf{v}^T\mathbf{x}\) is simply a polynomial in \(x\). Let us interpret the sum weighted by $p_i$ as computing an expectation value. The expectation value for the squared function $f(x)^2$ is positive 
\begin{equation}\label{DeltaProof}
    \bE [f(x)^2] \equiv \sum_i p_i f(x_i)^2 \geq 0.
\end{equation}
Since the vector \(\mathbf{v}\) is infinite dimensional, one can generate arbitrary polynomials in \(x\) with any degree. In a finite interval, one can approximate the delta function by an infinite sum of polynomials. In particular, on the interval ${-}1\leq x\leq1$ one can choose $f^2(x)$ to be the Landau Kernal $L_n(x)$ with $n$ being a large even integer, and
\eq
L_n(x)=(1-x^2)^n \frac{(2n)!(2n{+}1)}{(n!)^22^{2n{+}1}}\,.
\eqe
Choosing \(\mathbf{v}\) appropriately, we see that eq.(\ref{DeltaProof}) ensures with a positive $p_i $. Since the convex hull is by nature projective, we can rescale $x\rightarrow R x$ by any positive number $R$ and the geometry is unchanged. Therefore, this argument holds for any bounded interval of \(x\). A more formal treatment is required for \(R\rightarrow\infty\), given by~\cite{Moment1}.

\subsubsection*{Extra conditions for restricted moments}
If we further consider that \(\{x_i\}\) are restricted on a closed interval \([\alpha,\beta]\), this implies 
\begin{equation}
\label{ab_interval}
    \bE \left[f(x)^2 (x{-}\alpha)(\beta{-}x)\right] \geq 0
\end{equation}
since it gives us a positive sum of polynomial squared. Assuming eq.(\ref{ab_interval}), we can once again let \(f(x)^2\) approach delta function, enforcing \(p_i(x_i{-}\alpha)(\beta{-}x_i) \geq 0.\) Since \(p_i\) is positive, this guarantees \((x_i{-}\alpha)(\beta{-}x_i) \geq 0\), so \(x_i \in [\alpha, \beta]\). Rewriting eq.(\ref{ab_interval}) as 
\begin{equation}\label{ab_interval_matrix}
    \bE \left[f(x)^2 (x{-}\alpha)(\beta{-}x)\right]=\mathbf{v}^T \left( \sum_{i} p_{i}(x_i{-}\alpha)(\beta{-}x_i)\mathbf{x}_i\mathbf{x}_i^T \right) \mathbf{v} \geq 0\,,
\end{equation}
the matrix sandwiched by the arbitrary vector $\mathbf{v}$ must now be PSD. The new matrix is in fact a linear combination of the Hankel matrix and its shifted ones
\begin{align}
\label{hankel_abcd}
    \sum_{i} p_{i}(x_i{-}\alpha)(\beta{-}x_i) \left(\begin{array}{c}
        1 \\ x_i \\ x_i^2\\ \vdots
    \end{array}\right)
    \left(\begin{array}{cccc}
    1 & x_i & x_i^2 & \cdots
    \end{array}
    \right)    =  \begin{pmatrix}
    (\alpha{+}\beta)a_1{-}a_2{-}\alpha \beta a_0 & (\alpha{+}\beta)a_2{-}a_3{-}\alpha \beta a_1 & \cdots\\
    (\alpha{+}\beta)a_2{-}a_3{-}\alpha \beta a_1 & (\alpha{+}\beta)a_3{-}a_4{-}\alpha \beta a_2 & \cdots\\
    \vdots & \vdots & \ddots
    \end{pmatrix} \geq 0\,.
\end{align}
Therefore, the necessary and sufficient condition to have $x\in[\alpha,\beta]$ is given by the matrices eq.(\ref{aab}) and eq.(\ref{hankel_abcd}) being PSD. We can now apply this result to the following problems.
\

\subsubsection*{Half moment: $x\ge0$} 
     This simply corresponds to setting \(\alpha=0, \beta \rightarrow \infty\), giving
    \begin{equation}
        \bE \left[ f(x)^2 x \right] \geq 0 \Rightarrow
        \begin{pmatrix}
        a_1 & a_2 & a_3 & \cdots \\
        a_2 & a_3 & a_4 & \cdots \\
        a_3 & a_4 & a_5 & \cdots \\
        \vdots & \vdots & \vdots & \ddots
        \end{pmatrix}
        \geq 0,
    \end{equation}

\subsubsection*{Bounded moment: $x\in[0,1]$}

       In addition to \(x\geq0\), i.e., the positivity of the shifted Hankel, one also needs to impose \(x\leq1\), by setting  \(\alpha = 0, \beta = 1\)
    \begin{equation}\label{twisted_Hankel}
        \bE [(1-x) f(x)^2]\geq 0 \Rightarrow 
        \begin{pmatrix}
        (a_0-a_1) & (a_1-a_2) & (a_2-a_3) & \cdots \\
        (a_1-a_2) & (a_2-a_3) & (a_3-a_4) & \cdots \\
        (a_2-a_3) & (a_3-a_4) & (a_4-a_5) & \cdots \\
        \vdots & \vdots & \vdots & \ddots
        \end{pmatrix}
        \geq 0
    \end{equation}

\subsubsection*{Discrete moment: \( x \in \{s_i\}\)}

   Let \(s_1, s_2, ...,s_n\) be monotonically increasing. The previous constraints correspond to excluding regions \((-\infty,a)\) and \((b,\infty)\). If we instead exclude \(x\in(s_i, s_{i+1})\) by imposing \(\bE\left[f(x)^2(x-s_i)(x-s_{i+1})\right]\geq0\) and then apply this for every adjacent pair, it forces \(x\in\{s_i\}\) and gives us the discrete geometry.
    \begin{align}
    \label{discrete_matrix}
    \begin{split}
    (a) \quad &\bE \left[ f(x)^2(x-s_1) \right] \geq 0\Rightarrow
        	\begin{pmatrix}
       	 	(a_1-s_1a_0) & (a_2-s_1a_1) & \cdots\\
        	(a_2-s_1a_1) & (a_3-s_1a_2) & \cdots\\
        	\vdots & \vdots & \ddots
       	    \end{pmatrix} \geq 0. \\
    (b) \quad &\bE \left[ f(x)^2(x-s_{i})(x-s_{i+1})\right]  \cdots \geq 0 \\
       		 \Rightarrow 
       		 & \begin{pmatrix}
      		  (a_2-(s_{i}+s_{i+1})a_1+s_{i} s_{i+1} a_0) & (a_3-(s_{i}+s_{i+1})a_2+s_{i} s_{i+1} a_1) & \cdots\\
       		 (a_3-(s_{i}+s_{i+1})a_2+s_{i} s_{i+1} a_1) & (a_4-(s_{i}+s_{i+1})a_3+s_{i} s_{i+1} a_2) & \cdots\\
       		 \vdots & \vdots & \ddots
       		 \end{pmatrix}\cdots \geq 0. \\
    (c) \quad &\bE \left[f(x)^2(s_n-x)\right]\geq 0 \Rightarrow
        	\begin{pmatrix}
       		 (-a_1+s_n a_0) & (-a_2+s_n a_1) & \cdots\\
       		 (-a_2+s_n a_1) & (-a_3+s_n a_2) & \cdots\\
       		 \vdots & \vdots & \ddots
        	\end{pmatrix}\geq 0.
    \end{split}
    \end{align}

    Let us consider the first element of the matrix of (b) in eq.(\ref{discrete_matrix}), which turns out to be the boundary of cyclic polytope described in  \cite{Arkani-Hamed:2020blm}, up to a positive factor
    \begin{equation*}
        a_2-(s_i+s_{i+1})a_1+s_{i} s_{i+1} a_0=\frac{1}{(s_{i+1}-s_{i})}\,\textrm{det}
        \begin{pmatrix}
        a_0 & 1 & 1\\
        a_1 & s_i & s_{i+1}\\
        a_2 & s_i^2 & s_{i+1}^2
        \end{pmatrix}\,.
    \end{equation*}
    Since the matrix in $(b)$ of eq.(\ref{discrete_matrix}) is PSD, the above is positive and is exactly the \(\mathbb{P}^2\) polytopal constraint. Moving on to the larger leading principal minors, we find them having the form of the usual Hankel matrix with elements being the polytopal wall dotted into \(\vec{a}\) up to a positive factor.

Note the above constraints contain non-linear inequalities from the Hankel matrices. This is not what we expect for a cyclic polytope, which we know must only have linear boundaries. This is not a contradiction, as the above conditions are sufficient only in the infinite limit. Indeed one can check that the Hankel conditions converge to the polytope as we increase the size of the matrices considered.

\subsubsection{Double moment}\label{infconstr}
The  solution to the double moment problem, defined by
\eq\label{eqdouble}
a_{k,q}=\sum_i p_i x_i^k y_i^q, \quad p_i\ge 0
\eqe
is very similar in the infinite dimensional limit. However, so far only cases where both moments have identical support, such as $[0,\infty)\times[0,\infty)$ or $[0,1]\times[0,1]$, have been fully worked out in the literature, and are reviewed in Ref. \cite{Moment}.\footnote{The $K$ moment problem deals with more general moments, but only applies to compact domains \cite{article}.} In all cases however, the new central object in the solution is a generalization of the Hankel matrix, called a moment matrix, to be defined below. This result can be reproduced by extending the arguments in the previous section. This will allow us to derive constraints which are suited to our specific physical problem, which involves one moment defined on $x\in [0,1]$, and one discrete $y\in \{s_1,s_2,\ldots\}$. 

A product of two moment curves requires a complete basis of polynomials in \(x\) and \(y\), now given by \(
    \{ 1, x, y, x^2, xy, y^2, ... \}
\), to generate arbitrary function \(f(x,y)\). Therefore, we can assemble the couplings into the following matrix generalizing the Hankel matrix
\begin{equation}\label{product_Hankel}
    M(a) = \sum_{i} p_{i} \left(\begin{array}{c}
        1 \\ x_i \\ y_i \\ x_i^2 \\ x_i y_i \\ y_i^2 \\ \vdots
    \end{array}\right)
    \left(\begin{array}{ccccccc}
    1 & x_i & y_i & x^2_i & x_i y_i & y^2_i & \cdots
    \end{array}
    \right)
    =
    \begin{pmatrix}
        a_{0,0} & a_{1,0} & a_{0,1} & \cdots \\
        a_{1,0} & a_{2,0} & a_{1,1} & \cdots\\
        a_{0,1} & a_{1,1} & a_{0,2} & \cdots\\
        \vdots & \vdots & \vdots & \ddots
    \end{pmatrix}\,,
\end{equation}
which is known as the moment matrix. This particular arrangement will also be well suited to studying the truncated moment problem, where one is only interested to describe the space up to some order in $k$ and $q$.

The constraints for \(a_{k,q}\) to satisfy eq.(\ref{eqdouble}) are simply $M(a)\ge 0$. For example, we have, up to \(k,q \leq 2\),
\begin{equation}
    \begin{vmatrix}
        a_{0,0} & a_{1,0} \\
        a_{1,0} & a_{2,0} \\
    \end{vmatrix}, 
    \begin{vmatrix}
        a_{0,0} & a_{0,1} \\
        a_{0,1} & a_{0,2} \\
    \end{vmatrix},
    \begin{vmatrix}
        a_{0,0} & a_{1,1} \\
        a_{1,1} & a_{2,2} \\        
    \end{vmatrix},
    \begin{vmatrix}
        a_{0,0} & a_{1,0} & a_{0,1}\\
        a_{1,0} & a_{2,0} & a_{1,1}\\
        a_{0,1} & a_{1,1} & a_{0,2}
    \end{vmatrix}
    \geq
    0.\nn
\end{equation}
We note that conditions such as the last two were not previously considered in~\cite{Arkani-Hamed:2020blm}, and are also related to an inequality used in~\cite{Caron-Huot:2020cmc} to analytically obtain a bound we will also aim to study from the geometric perspective.  

Similarly, the additional constraints for restricted \(x,y\) can be straight forwardly derived. One imposes 
\[\bE \left[f(x,y)^2 (x-x_a)(x_b-x)\right] \geq 0\]
to force \(x\in[x_a,x_b]\) and imposes 
\[\bE \left[f(x,y)^2 (y-y_a)(y-y_b)\right] \geq 0\]
to exclude \(y\in(y_a,y_b)\). So for each case we have the following extra constraints:
\begin{enumerate}
    \item \textbf{Half moment} 
    
    For \(x \in [0,\infty)\), we require
    \begin{eqnarray}
        & \bE \left[f(x,y)^2x\right] \geq 0 \nn \\
        \Rightarrow  & \sum_{i,j} p_{i,j}x_i \left(\begin{array}{c}
            1 \\ x_i \\ y_i\\ \vdots
        \end{array}\right)
        \left(\begin{array}{cccc}
        1 & x_i & y_i & \cdots
        \end{array}
        \right)
        =
        \begin{pmatrix}
        a_{1,0} & a_{2,0} & a_{1,1} & \cdots \\
        a_{2,0} & a_{3,0} & a_{2,1} & \cdots\\
        a_{1,1} & a_{2,1} & a_{1,2} & \cdots\\
        \vdots & \vdots & \vdots & \ddots
        \end{pmatrix} \geq 0
    \end{eqnarray}
    This is the two-dimensional generalization of the shifted Hankel matrix. In short,
    \begin{equation}\label{shift_x}
        M^{\text{shift},x}(a) \equiv M(a)|_{a_{k,q}\rightarrow  a_{k+1,q}} \geq 0.
    \end{equation}
    If \(y\) is also required to be positive, one has to impose
    \begin{equation}
        M^{\textrm{shift},y}(a) \equiv M(a)|_{a_{k,q}\rightarrow a_{k,q+1}} \geq 0.
    \end{equation}
    \item \textbf{Bounded moment} 
    
   For \(x\in[0,1]\), then one has to impose, besides eq.(\ref{shift_x}), positivity on the twisted Hankel matrix
    \begin{equation}
        \bE \left[f(x,y)^2(1-x)\right] \geq 0
        \Rightarrow \begin{pmatrix}
        (a_{0,0}-a_{1,0}) & (a_{1,0}-a_{2,0}) & (a_{0,1}-a_{1,1}) & \cdots\\
        (a_{1,0}-a_{2,0}) & (a_{2,0}-a_{3,0}) & (a_{1,1}-a_{2,1}) & \cdots\\
        (a_{0,1}-a_{1,1}) & (a_{1,1}-a_{2,1}) & (a_{0,2}-a_{1,2}) & \cdots\\
        \vdots & \vdots & \vdots & \ddots
        \end{pmatrix} \geq 0.
    \end{equation}
    This is the generalized version of eq.(\ref{twisted_Hankel}).
    \item \textbf{Discrete moment}
    
    Suppose \(y\) is discrete and can only take values \(\{s_i\}\), then
    \begin{equation}
    \label{bi-variate_discrete}
        \bE \left[ f(x,y)^2 (y-s_1) \right],\ \bE \left[ f(x,y)^2 (y-s_{i})(y-s_{i+1}) \right], \bE \left[ f(x,y)^2 (s_n - y) \right] \geq 0.
    \end{equation}
    In other words, 
    \begin{align}
    \label{2d_discrete_matrix}
    \begin{split}
    (a) \quad &\bE \left[ f(x,y)^2 (y-s_1) \right] \geq 0\Rightarrow
        	\begin{pmatrix}
       	 	(a_{0,1}-s_1a_{0,0}) & (a_{1,1}-s_1 a_{1,0}) & \cdots\\
        	(a_{1,1}-s_1 a_{1,0}) & (a_{2,1}-s_1a_{2,0}) & \cdots\\
        	\vdots & \vdots & \ddots
       	    \end{pmatrix} \geq 0. \\
    (b) \quad &\bE \left[ f(x,y)^2 (y-s_{i})(y-s_{i+1})\right]  \cdots \geq 0 \\
       		 \Rightarrow 
       		 & \begin{pmatrix}
      		  (a_{0,2}-(s_{i}+s_{i+1})a_{0,1}+s_{i} s_{i+1} a_{0,0}) & (a_{1,2}-(s_{i}+s_{i+1})a_{1,1}+s_{i} s_{i+1} a_{1,0}) & \cdots\\
       		 (a_{1,2}-(s_{i}+s_{i+1})a_{1,1}+s_{i} s_{i+1} a_{1,0}) & (a_{2,2}-(s_{i}+s_{i+1})a_{2,1}+s_{i} s_{i+1} a_{2,0}) & \cdots\\
       		 \vdots & \vdots & \ddots
       		 \end{pmatrix}\cdots \geq 0. \\
    (c) \quad &\bE \left[f(x,y)^2 (s_n-y)\right]\geq 0 \Rightarrow
        	\begin{pmatrix}
       		 (-a_{0,1}+s_n a_{0,0}) & (-a_{1,1}+s_n a_{1,0}) & \cdots\\
       		 (-a_{1,1}+s_n a_{1,0}) & (-a_{2,1}+s_n a_{2,0}) & \cdots\\
       		 \vdots & \vdots & \ddots
        	\end{pmatrix}\geq 0.
    \end{split}
    \end{align}
    
    The first element of the matrix (b) above being positive
    \begin{equation*}
    	a_{0,2}-(s_{i}+s_{i+1})a_{0,1}+s_{i} s_{i+1} a_{0,0}
    	=
    	\frac{1}{(s_{i+1}-s_{i})}\,\textrm{det}
        \begin{pmatrix}
        a_{0,0} & 1 & 1\\
        a_{0,1} & s_i & s_{i+1}\\
        a_{0,2} & s_i^2 & s_{i+1}^2
        \end{pmatrix}\ge 0\,, 
    \end{equation*}
    is exactly the \(\mathbb{P}^2\) polytopal wall, generalized to the bi-variate case. The constraints in eq.(\ref{2d_discrete_matrix}) should give us the discrete-y geometry in the infinite dimension limit.     
\end{enumerate}

We conjecture that in the infinite dimension limit, choosing the appropriate conditions associated with the two moments from above, yields the complete sufficient conditions.    
\subsection{Truncated moment problem}
\paragraph{Extension theorems for single moment}
Above we have seen that an infinite dimensional space can be carved out by imposing positivity of particular infinite dimensional matrices. However, suppose we are interested only in a finite dimensional space, $a=\{a_0,a_1,\ldots,a_{2n}\}$. How well can we describe this space by using only a finite number of constraints? 

An obvious practical approach is to impose positivity of matrices of increasing but finite size $m$. Clearly we expect $m$ to be at least equal to $n$, in order to constrain all dimensions, but it is less obvious that is sufficient.  One could also expect this approach will only converge to the actual solution in the infinite limit, leading to discrepancies whenever $m$ is finite. Surprisingly, we will see that the correct answer is we only need $m=2n+1$.

Let us consider some simple examples. Say we are interested in the space $a_0,a_1,a_2$, $x\in \mathbb{R}$. The simplest guess is we should require the 2$\times2$ Hankel matrix to be positive:
\eq
H_2=\begin{pmatrix}
a_0&a_1\\
a_1&a_2
\end{pmatrix}\ge 0\quad  \Rightarrow \quad a_0\ge 0\ \textrm{and } a_0a_2-a_1^2\ge0
\eqe
However, these are actually not the complete conditions. The singular geometry with coordinate $a_0=0$ is not correctly described. Plugging $a_0=0$ into our conditions, we obtain $a_1=0$, but we find no constraint on $a_2$. This is clearly not correct, as $a_0=0\Rightarrow p_i=0$, so we should have obtained $a_2=0$. The missing conditions are known as the flat or positive extensions \cite{Curto}, and involve adding extra moments to construct larger Hankel matrices. In our case, we need to extend the space to $\{a_0,a_1,a_2,{\bf a_3},{\bf a_4}\}$. We can write a larger Hankel matrix
\eq
H_3=\begin{pmatrix}
a_0&a_1&a_2\\
a_1&a_2&{\bf a_3}\\
a_2&{\bf a_3}&{\bf a_4}
\end{pmatrix}
\eqe
Now there are two equivalent conditions we can impose: 
\begin{itemize}
\item The extension is flat: $\textrm{Rank}[H_3]=\textrm{Rank}[H_2]$, $\forall a_3,a_4$
\item The extension is positive: $H_3\ge 0$, $\forall a_3,a_4$

\end{itemize}
Imposing either we find $a_2=0$, as expected. For a slightly less trivial example, consider the sequence $a_k=\{1,1,1,1,a_4\}$, and we wish to find what values of $a_4$ ensure $a_k$ is a moment sequence. The Hankel matrix
\eq
H_3=\begin{pmatrix}
1&1&1\\
1&1&1\\
1&1&a_4\end{pmatrix}
\eqe
is PSD for $a_4\ge 1$. However, this is not sufficient. Because the $2\times2$ submatrix is actually singular, that requires there is just one state, with $p_1=1,x_1=1$, implying in fact that  $a_4=1$. The extension theorem states that the full conditions are, in terms of the extended sequence $a_k=(1,1,1,a_4,{\bf a_5},{\bf a_6})$, the larger Hankel matrix
\eq
H_4=\begin{pmatrix}
1&1&1&1\\
1&1&1&a_4\\
1&1&a_4&{\bf a_5}\\
1&a_4&{\bf a_5}&{\bf a_6}\end{pmatrix}
\eqe
must be PSD, for all $a_5,a_6$. The positive extension requires $\textrm{Det}[H_4]\ge 0$, which implies $a_4\le 1$. Together with the original condition $\textrm{Det}[H_3]\ge 0$ , we obtain $a_4=1$. Equivalently, the ``flat extension" condition is the requirement that $\textrm{Rank}[H_4]=\textrm{Rank}[H_3]$. This latter condition similarly requires $a_4=1$. 

This subtle issue will have even more important consequences for the bi-variate problem, where such discrepancies extend beyond singular points of the geometry to affect non-trivial regions of the allowed space. 

\paragraph{Extension conjectures for bi-variate problem}

The solution to the truncated bi-variate problem is not as established, but an equivalent statement is conjectured to hold \cite{truncateconj}. The best we can do therefore is impose the infinite solution constraints derived in Section \ref{infconstr}, up to some finite order. That is, if we are interested in some finite dimensional space of moments up to order $m$, we must impose constraints involving the extended sequence up to higher order $m'\ge m$. As we increase the number of orders considered, this approximate solution must converge to the exact solution. In all cases studied in Sections \ref{sec3} and \ref{sec4} we find this approximate solution quickly converges to numerical results, already when considering just constraints at order $m'=m+1$.

To be concrete about the constraints we will use in our specific physical problem, let us assume we are interested in the finite dimensional space given by moments 
\eq
a_{k,q}=\sum_i p_i x_i^k y_i^q, \quad x\in[0,1], \ y\in \{s_i\}
\eqe 
with $k\ge 2$, $k-q\ge 2$ and for some $k\le m$. First, we re-organize the moment matrix in a form more appropriate to these requirements. Instead of moments $a_{k,q}\sim x^k y^q$, we rewrite $a_{k,q}\sim x^2 x^k (x y)^q$. Then the moment matrix becomes
\eq
M_{2n+2}(a_{k,q})= \begin{blockarray}{ccccccccc}
&1 & x & (xy) & x^2 & x(xy)&(xy)^2& \cdots&(xy)^n\\
\begin{block}{c(cccccccc)}
1&a_{2,0}& a_{3,0} &a_{3,1}&a_{4,0}&a_{4,1}&a_{4,2}&\cdots&a_{n+2,n}\\
x&a_{3,0}&a_{4,0}&a_{4,1}&a_{5,0}&a_{5,1}&a_{5,2}&&\\
(xy)& a_{3,1}&a_{4,1}&a_{4,2}&a_{5,1}&a_{5,2}&a_{5,3}&&\\
x^2&a_{4,0}& a_{5,0} &a_{5,1}&a_{6,0}&a_{6,1}&a_{6,2}&&\vdots\\
x(x y) &a_{4,1}& a_{5,1} &a_{5,2}&a_{6,1}&a_{6,2}&a_{6,3}&&\\
(xy)^2 &a_{4,2}& a_{5,2} &a_{5,3}&a_{6,2}&a_{6,3}&a_{6,4}&&\\
\vdots &\vdots&& &&&&\ddots& \\
(xy)^n &a_{n+2,n}&&&\cdots&&&&a_{2n+2,2n}\\
\end{block}
\end{blockarray}
\eqe
In terms of this matrix we will impose the following constraints to obtain an approximate solution for $a_{k,q}$ with $k\le m$
\eqa\label{finaleq}
\nonumber &&M_{[m]}(a_{k,q})\ge 0\\
\nonumber &x\ge 0:&M_{[m]}(a_{k+1,q})\ge 0\\
\nonumber &x\le1:& M_{[m]}(a_{k,q}-a_{k+1,q})\ge0\\
\nonumber &xy\ge 0:& M_{[m]}(a_{k+1,q+1}) \geq 0\\
\nonumber &y\in \{s_i\}:& P^{(i)}_{k\le m+1,q}\ge 0\\
&y\in \{s_i\}:&M_{[m]} (P^{(i)}_{k,q})\ge 0
\eqae
where $[m]=m$ for even and $[m]=m+1$ for odd $m$, and 
\eq
P_{k,q}^{(i)}=s_i s_{i+1}a_{k,q}-(s_i+s_{i+1})a_{k,q+1}+a_{k,q+2}\,.
\eqe
One may worry that $P^{(i)} $ still contains an infinite number of constraints, since we must allow the possibility for an infinite number of spins $\{s_i\}$. However, as we discuss in the next section, the null planes intersect the polytope at a finite number of locations. The roots of the null constraint will directly inform us which are the relevant polytope walls that we need to consider.

\subsection{Symmetry planes}
An important ingredient in the geometry will be the symmetry planes, which reflect some symmetry associated to the ordering of external particles in the 4 point amplitude. The symmetry plane can generically be described by any linear function in $a_{k,q}$, $f(a_{k,q})=0$. From the UV point of view, the null constraints act by imposing non-trivial conditions on the spectrum. This reduces the allowed space of couplings, since now in couplings given by $a_{k,q}=\sum p_i x_i^k y_i^q$, $p_i$, $x_i$, and $y_i$ are  further restricted. From the geometry point of view, the symmetries reduce the allowed space by simply picking slices of the EFThedron. For now we will only use the purely geometric aspect of the null conditions, and return to the UV perspective in Section \ref{UV}.
 
\begin{figure}[h] 
   \centering
   \includegraphics[width=3.3in]{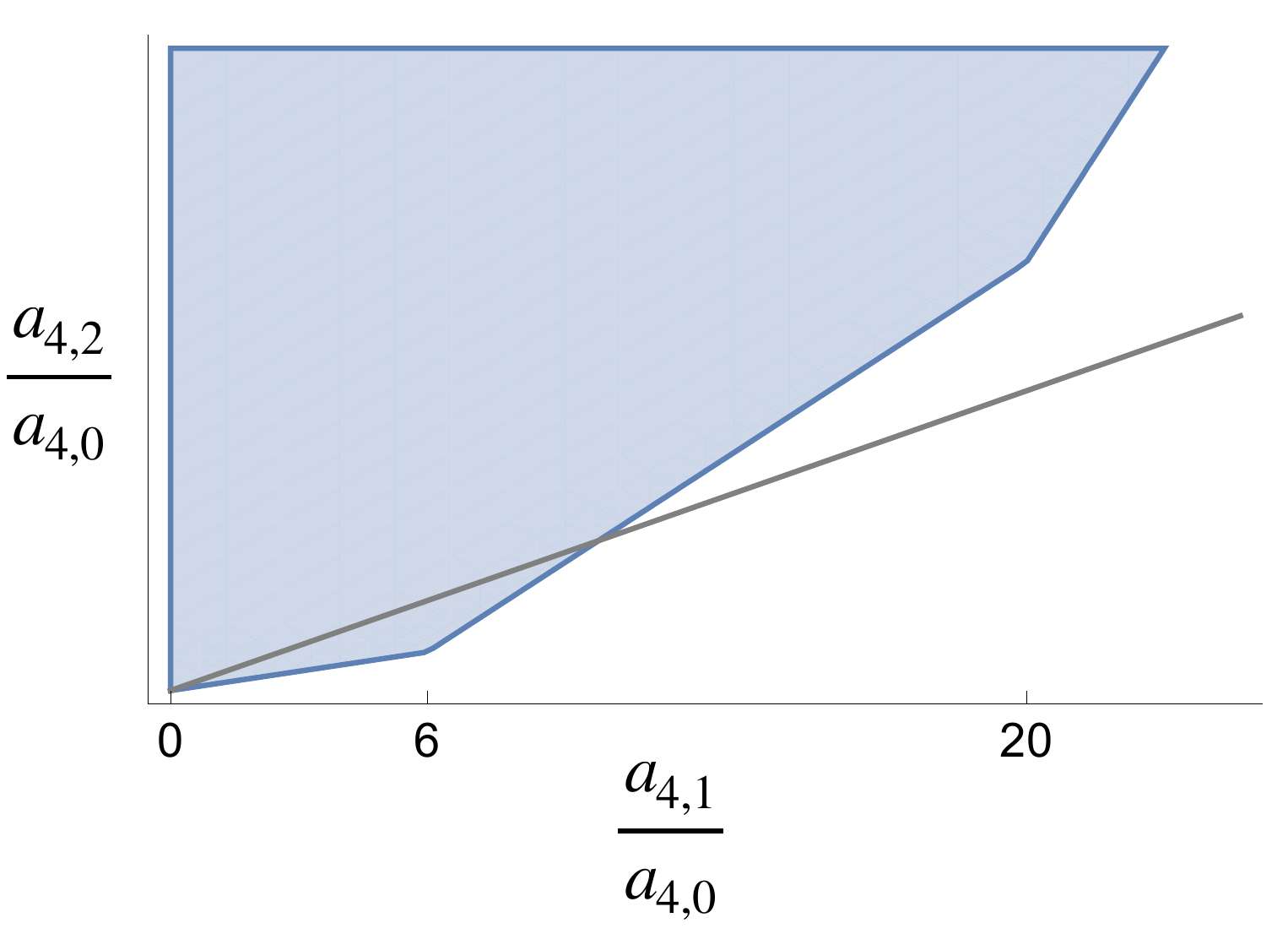} 
   \caption{1. $(a_{4,2}/a_{4,0},a_{4,1}/a_{4,0})$ polytope and $a_{4,2}=8a_{4,1}$ symmetry plane}
   \label{2dplot}
\end{figure}

 In general, a symmetry plane can combine couplings at both different $k$ and $q$. This is the case of the monodromy plane, which was considered in~\cite{Huang:2020nqy}. In our present discussion we will only consider cyclic and permutation invariance, so it will turn out each symmetry plane only acts at equal $k$, relating couplings at different $q$. While a plane acts in one particular $k$ level, it also induces non trivial effect to both higher and lower $k$ couplings. For instance, let us illustrate a simple example, that will also be relevant to the $su$-EFThedron. We will explain this in more detail in the following section, but for now we 
 just claim that permutation invariance, at order $k=4$, implies the couplings must be linearly related (briefly, this is due to the fact that there is just one permutation invariant kinematic object at $k=4$, namely $(s^2+t^2+u^2)^2$). In particular, we have
 \eq\label{nullctr}
a_{4,2}=8a_{4,1}
\eqe
The polytope boundaries at $k=4$ are $\langle \vec{a}_4,J_i^2,J_{i+1}^2\rangle\ge 0$, $J^2=\ell(\ell+1)$, where for $su$ $\ell\in\{0,2,4,6,\ldots\}$. The first few are
\eq\label{polyplane}
-6 a_{4,1} + a_{4,2}\ge 0,\quad 120 a_{4,0} - 26 a_{4,1} + a_{4,2}\ge 0,\quad 840 a_{4,0} - 62 a_{4,1} + a_{4,2}\ge 0
\eqe
We obtain the intersection in Figure \ref{2dplot}.
It is clear the symmetry plane intersects only one of the polytope walls, namely the one defined by vertices $J^2_i=6$, $J^2_{i+1}=20$. This can be directly explained by looking at the dispersive form of the null constraint in eq.(\ref{nullctr})
\eq
n_4=\sum p_i \frac{1}{m_i^{10}} J_i^2\left(J_i^2-8\right)
\eqe
We observe the summand has a root at $J^2=8$, so the relevant polytope wall is the one with vertices on either side of this root, namely $J^2_i=6$, $J^2_{i+1}=20$ in our $su$ case when there is no $J_i^2=12$.

\begin{figure}[h] 
   \centering
   \includegraphics[width=2.7in]{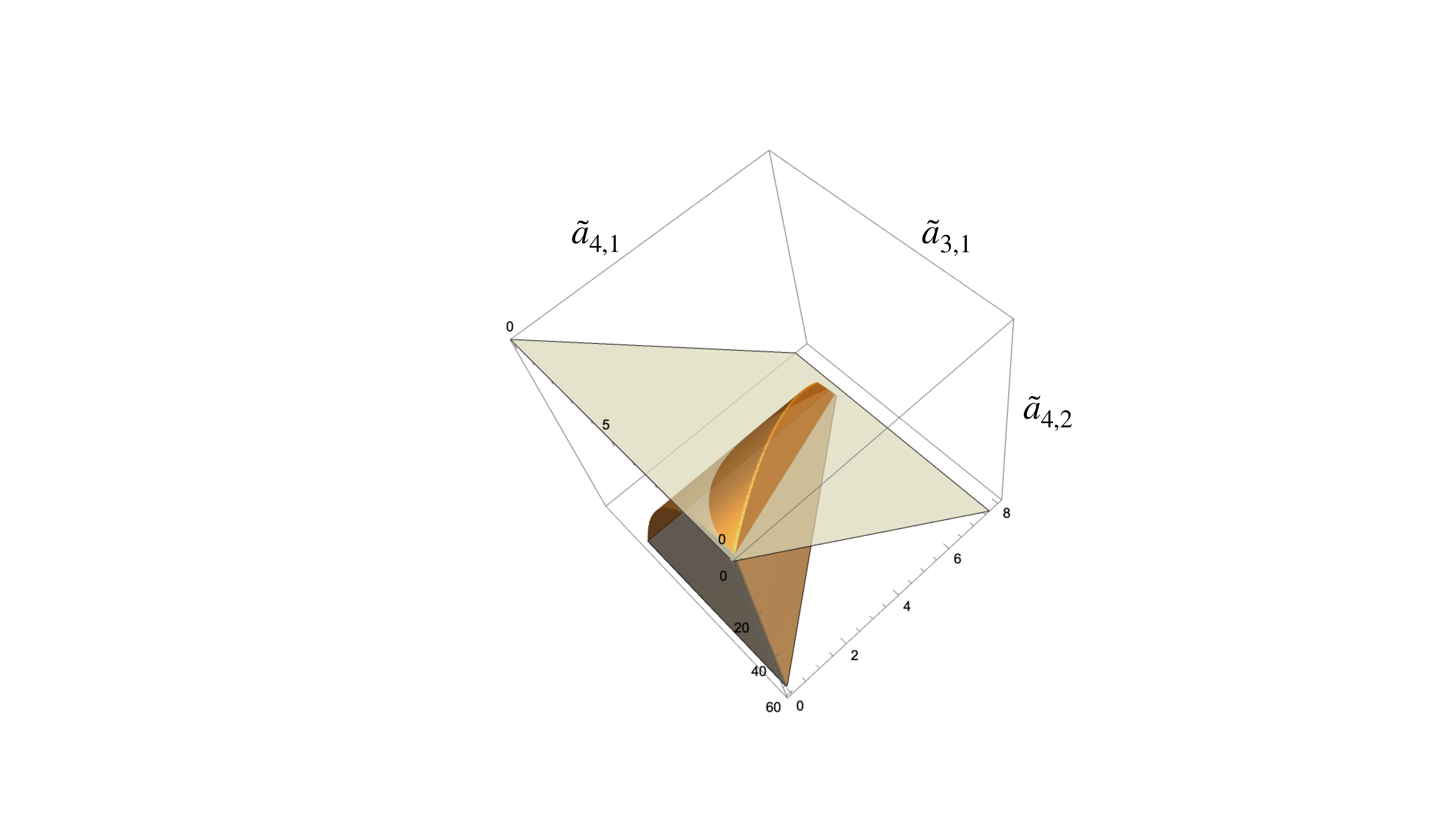} 
   \includegraphics[width=2.75in]{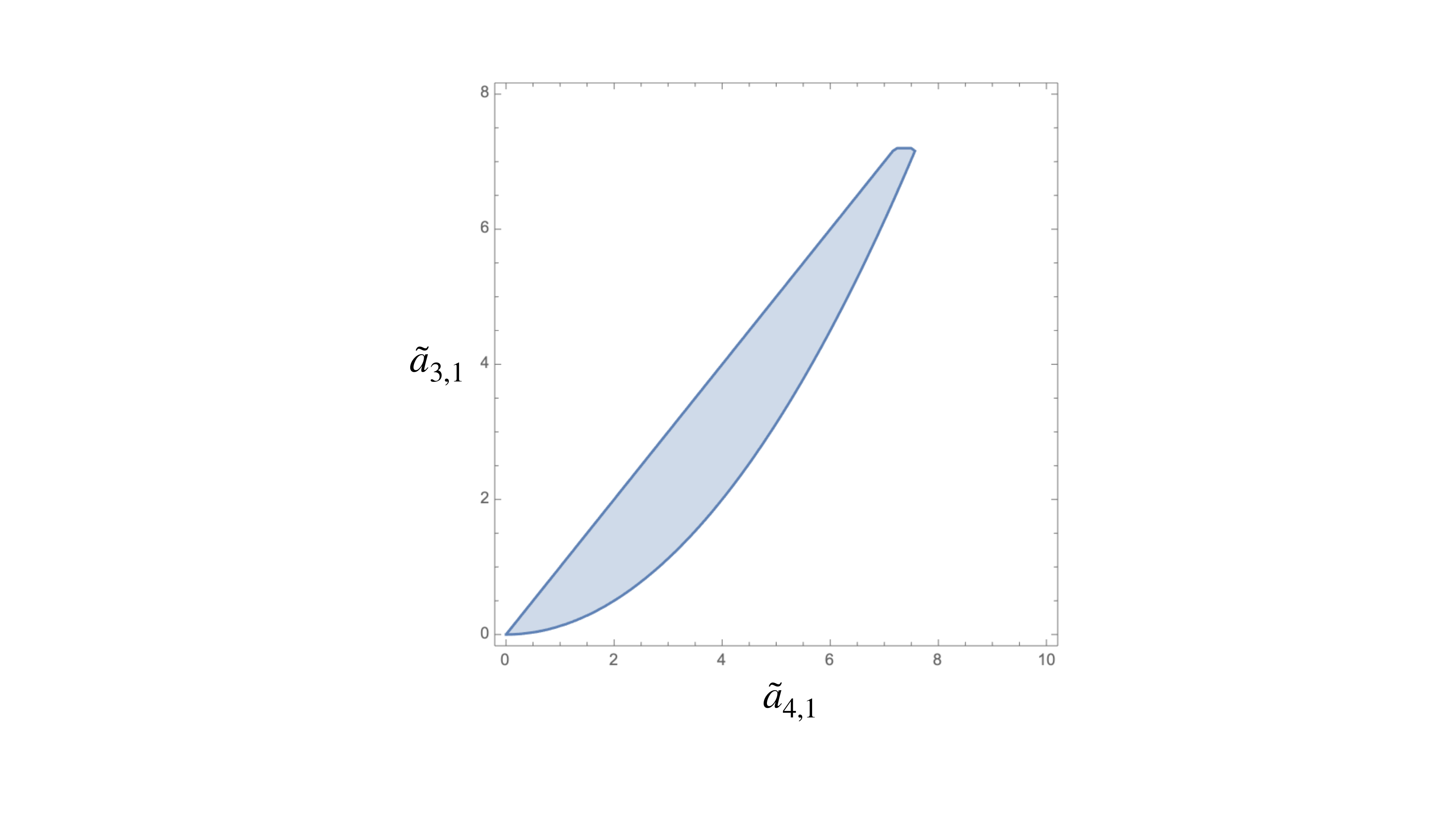} 
   \caption{$(\tilde{a}_{3,1},\tilde{a}_{4,1},\tilde{a}_{4,2})$ space intersected with $a_{4,2}=8a_{4,1}$ symmetry plane, and the resulting $(\tilde{a}_{3,1},\tilde{a}_{4,1})$ slice}
   \label{3dplot}
\end{figure}

For a 3D example, we can consider the geometry of  $(\tilde{a}_{3,1},\tilde{a}_{4,1},\tilde{a}_{4,2})$, where $\tilde{a}_{k,q}$ is just the projective coordinate  $\frac{a_{k,q}}{a_{2,0}}$. This scape is carved out by the constraints derived in the previous section, including the polytope wall $\langle \vec{a}_4,6,20\rangle$. In Figure \ref{3dplot} we plot the intersection with the symmetry plane, now a finite region.

\section{The $s$-EFThedron}\label{sec3}
As a warm-up for latter analysis, let's consider a toy-model EFT where the light states are adjoint or bi-fundamental states that admit an ordering. That is, we can decompose the amplitude of the massless scalars in terms of color traces and focus on the single trace amplitudes
\eq
M(\phi^a\phi^b\phi^c\phi^d)tr(T^aT^bT^cT^d)
\eqe 
Due to its trace structure, at fixed $t$, the UV completion will only involve $s$-channel discontinuities. Thus the dispersive representation of the EFT coefficients will be given as
\eq\label{DispS}
\sum_{k,q} g^{(s)}_{k,q} s^{k-q} t^q=-\sum_i p_{i}\;\frac{P_{\ell_i}(1{+}2t/m_i^2)}{s{-}m^2_i},\quad p_i>0
\eqe
where $P_\ell(\cos\theta)$ is the Legendre polynomial and the equality is understood in terms of Taylor expansion in $s,t$ keeping terms with $k{-}q\geq2$. The latter constraint is required for the absence of subtraction terms in the dispersion relation (see~\cite{Arkani-Hamed:2020blm} for a general discussion). In the following, we will study the geometry emerging for these dispersion relations and their implications for the couplings. 

\subsection{The $s$-EFThedron as a rotation}
Expanding the Legendre polynomial
\eq
P_{\ell}(1+2\delta)=\sum_{q=0}^{\ell} v_{\ell,q} \delta^q\,,
\eqe
the EFT couplings $g_{k,q}$ through the dispersive representation are therefore given by 
\eq
\left(\begin{array}{cccc}g^{(s)}_{2,0} & \; & \; & \; \\ g^{(s)}_{3,0} & g^{(s)}_{3,1} & \; & \; \\ g^{(s)}_{4,0} & g^{(s)}_{4,1} & g^{(s)}_{4,2} & \; \\ \vdots & \vdots & \vdots &  \ddots \end{array}\right)= \sum_i \frac{p_i}{m^6_i} \left(\begin{array}{c} 1 \\  \frac{1}{m^2_i} \\  \frac{1}{m^4_i} \\ \vdots\end{array}\right)\otimes (v_{\ell_i,0}, v_{\ell_i,1}, v_{\ell_i,2},\cdots)\,.
\eqe
where $v_{\ell,q}$, in terms of $J^2=\ell(\ell+1)$, is
\eqa
v_{\ell,q}=\frac{\prod_{a=1}^q(J^2{-}a(a{-}1))}{(q!)^2}\,.
\eqae
Importantly, as $v_{\ell,q}$ is a complete polynomial of degree $q$ in $J^2$, we can invert the relation via a GL transformation, and so we can write
\eq
{\bf{G}}_{J,v}\cdot \vec{v}=\vec{J^2}\, .
\eqe
For example, the $5\times5$ rotation is given by
\eq\label{ExplicitGL}
{\bf{G}}_{J,v}=\begin{pmatrix}
 1 & 0 & 0 & 0 & 0 \\
 0 & 1 & 0 & 0 & 0 \\
 0 & 2 & 4 & 0 & 0 \\
 0 & 4 & 32 & 36 & 0\\
 0 & 8 & 208 & 720 & 576
\end{pmatrix}
\eqe
Writing $\vec{g}^{\;(s)}_{k}=(g^{(s)}_{k,0},g^{(s)}_{k,1},\cdots)$, we define the new couplings $a_{k,q}$ via the same GL transformation
\eq
 \vec{a}_{k}={\bf{G}}_{J,v}\cdot \vec{g}^{\,(s)}_{k}\,,
\eqe
where now 
\eq
\left(\begin{array}{cccc}a_{2,0} & \; & \; & \; \\ a_{3,0} & a_{3,1} & \; & \; \\ a_{4,0} & a_{4,1} & a_{4,2} & \; \\ \vdots & \vdots & \vdots &  \ddots \end{array}\right)= \sum_i \frac{p_i}{m^6_i} \left(\begin{array}{c} 1 \\  \frac{1}{m^2_i} \\  \frac{1}{m^4_i} \\  \vdots\end{array}\right)\otimes (1, J^2_i, J_i^4,\cdots)\,.
\eqe
Thus we see that the dispersion representation in eq.(\ref{DispS}) translates to the constraint that the EFT couplings $g^{(s)}_{k,q}$ must live in the space of product moment curves. Moreover, one of the moments will be discrete since $J^2$ can only take discrete values, $\{0,2,6,12,20,\cdots\}$, while the other will only take values in the interval $[0,1]$, as we see next.

For a EFT to be well defined, there must be a gap between the UV states and the low energy (massless) states. In other words we have that all $m_{i}\ge M>0$. These observations are easily incorporated by considering the rescaled couplings
\eq\label{Gap1}
\left(\begin{array}{c} a_{k,q} \\ a_{k{+}1,q} \\  a_{k{+}2,q} \\ \vdots \end{array}\right)=\left(\begin{array}{c} a_{k,q} \\ a_{k{+}1,q}M^2 \\  a_{k{+}2,q}M^4 \\ \vdots \end{array}\right)=\sum_{a} p_a\left(\begin{array}{c} 1 \\ \frac{M^2}{m^2_a} \\  \frac{M^4}{m^4_a}  \\ \vdots \end{array}\right)=\sum_{a} p_a\left(\begin{array}{c} 1 \\ x_a \\ x^2_a  \\ \vdots \end{array}\right)\,,
\eqe
where $x_a$ is bounded from above and below, $0<x_a\leq1$, so the results for bounded intervals derived in the previous section can be used. We will often normalize $M=1$ in the rest of the paper.

In conclusion, we have shown that linear combinations of couplings $g^{(s)}$ of the $s$-channel EFThderon must satisfy eqns.(\ref{finaleq}).

Finally, since the amplitude is defined with a color trace, we can choose the states in such a way that it is cyclic invariant, then we have
\eq
M(s,t)=M(t,s)
\eqe
simply implying $g_{k,q}=g_{k,k-q}$, for $k\ge 2$ and $k-q\ge 2$. Couplings satisfying the previous identities are said to live on the cyclic symmetry plane, and the theory space for the EFT couplings is simply the intersection of the cyclic plane with the product geometry. 

In summary, we have shown that the $s$-EFThedron is simply a rotation of the bi-variate moment problem defining the $a$-geometry, intersected with the cyclic symmetry plane!


\subsection{Explicit region analysis}

Having obtained the constraints, the allowed region for the couplings can be computed, and in this section, we will be interested in the \((g_{3,0},g_{3,1})\)-space. We first carve out the \(a\) space then rotate back to the \(g\) space. Since \(a_{k,0} = g_{k,0}\) and \(a_{k,1} = g_{k,1}\), mapping those couplings is trivial. We will consider constraints by order in $k$.
\begin{itemize}
\item  $k= 3$
The only conditions are
\eqa
\nonumber a_{2,0},a_{3,0},a_{3,1}\ge0\\
a_{2,0}-a_{3,0}\ge0
\eqae

\item $k=4$ 
We now have the moment matrix, and the shifted matrix, which in this case contains just $a_{4,1}$
\begin{equation}\label{s moment}
    \begin{pmatrix}
	a_{2,0} & a_{3,0} & a_{3,1}\\
	a_{3,0} & a_{4,0} & a_{4,1}\\
	a_{3,1} & a_{4,1} & a_{4,2}
	\end{pmatrix} \geq 0,\quad  a_{4,1}\ge 0
\end{equation}

Next, the twisted Hankel constraints are
\begin{equation}
	(a_{3,0} - a_{4,0}) , \, (a_{3,1} - a_{4,1}) \geq 0.
\end{equation}
Since for $s$-EFThedron \(J^2_i=\ell_i(\ell_i+1)\) can only take discrete values,  \(J^2_i=\{0,2,6,12,20,\cdots\}\), we only have the following polytope constraints
\begin{equation}
\langle \vec{a}_4,J^2_i,J^2_{i+1} \rangle \ge0,
\end{equation}
with $\vec{a}_4=(a_{4,0},a_{4,1},a_{4,2})$. Eliminating \(a_{4,0},a_{4,1},a_{4,2}\) in these inequalities, we obtain the same simple region carved out by the $k=3$ constraints alone
\begin{equation}
    1 \geq \tilde{a}_{3,0} \geq 0, \, \tilde{a}_{3,1} \geq 0.
\end{equation}
where we remind the notation $\tilde{a}_{k,q}\equiv \frac{a_{k,q}}{a_{2,0}}M^{2(k-2)}$.

As expected we see no improvement without the presence of a null constraint.

\item  \(k = 5\) 
The new constraints are
\begin{equation*}
    \begin{pmatrix}
    a_{3,0} & a_{4,0} & a_{4,1} \\
    a_{4,0} & a_{5,0} & a_{5,1} \\
    a_{4,1} & a_{5,1} & a_{5,2} \\
    \end{pmatrix}, \,
    \begin{pmatrix}
    a_{3,1} & a_{4,1} & a_{4,2} \\
    a_{4,1} & a_{5,1} & a_{5,2} \\
    a_{4,2} & a_{5,2} & a_{5,3} \\
    \end{pmatrix}
    \geq 0,
\end{equation*}
\begin{equation}\label{3x3s}
     \begin{pmatrix}
    a_{2,0}-a_{3,0} & a_{3,0}-a_{4,0} & a_{3,1}-a_{4,1} \\
    a_{3,0}-a_{4,0} & a_{4,0}-a_{5,0} & a_{4,1}-a_{5,1}\\
    a_{3,1}-a_{4,1} & a_{4,1}-a_{5,1} & a_{4,2}-a_{5,2} \\
    \end{pmatrix},\,
    \begin{pmatrix}
    a_{4,0}-a_{5,0}
    \end{pmatrix},\,
    \begin{pmatrix}
    a_{4,1}-a_{5,1}
    \end{pmatrix} \geq 0. 
    \end{equation}

Next, the $k=5$ polytope boundaries will be given by
\eqa
\nonumber \langle 0,\vec{a}_5,J^2_i,J^2_{i+1}\rangle \ge0\\
\langle \vec{a}_5,J^2_i,J^2_{i+1},\infty\rangle \ge0
\eqae
with $\vec{a}_5=(a_{5,0},a_{5,1},a_{5,2},a_{5,3})$. 

Crucially, at \(k=5\) the first null constraint \(g_{5,2}=g_{5,3}\) appears, implying 
\eq
a_{5,3}=17 a_{5,2}-30a_{5,1}
\eqe
Written in terms of UV states, the associated null constraint
\eq
n_5=\sum_i \frac{p_i}{m_i^{12}}\left(J_i^6-17 J_i^4+30J_i^2\right)=0
\eqe
has roots at $J^2=2,15$. This means the only relevant walls will be those with vertices on either side of $J^2=15$, namely $\langle 0,\vec{a}_5,12,20\rangle$ and $\langle \vec{a}_5,12,20,\infty\rangle$. 
\end{itemize}

\begin{figure}[h]
	\centering
	\includegraphics[width=0.75\textwidth]{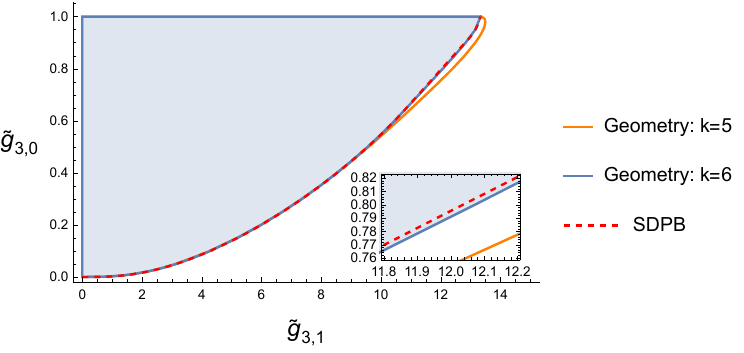}	
	\caption{Geometry bound and SDPB bound for \((\tilde{g}_{3,1},\tilde{g}_{3,0})\) space. Including higher order couplings and constraints improves the geometry bound. The region with the maximal mismatch is zoomed in for clarity.}
	\label{g3_k=5}
\end{figure}

The matrices in eqns.(\ref{3x3s}) make solving the inequalities difficult, so we will resort to a numerical mapping of the space. 
To solve for the \(g_3\) space, interior-point method was used to find the extremal value of \(\tilde{g}_{3,1}\) under the constraints listed above. By Fixing \(\tilde{g}_{3,0}\) to various different values, the space can be scanned over and we find a finite allowed region, which we compare with an SDPB analysis in Figure \ref{g3_k=5}. Extracting the upper bounds, we obtain
\eq
\tilde{g}_{3,1}^{(k=5)}\le 13.47
\eqe
from geometry compared to 
\eq
\tilde{g}_{3,1}\le 13.33
\eqe
from SDPB. Including higher order constraints involving couplings at order $k=6$ further improves our results, where the upper bound
\eq
\tilde{g}_{3,1}^{(k=6)}\le 13.33
\eqe
now matches exactly with the bound from SDPB. A small deviation can still be observed around \(\tilde{g}_{3,1} = 12\), which we expect is a consequence of the truncation at finite order. We see that the region converges rapidly as the truncation increases, and expect a convergence to the exact solution in the infinite-\(k\) limit. In the next section for the full EFThedron, where the symmetry plane intersects at lower order in $k$, we will demonstrate an analytic result exactly matching with results from SDPB in all regions. 

\subsubsection*{A comparison with previous bounds}
In the previous paper introducing the EFThedron \cite{Arkani-Hamed:2020blm}, a set of bounds of the s-channel geometry were given in terms of $g_{k,q}$, organized as follows: fixed $q$ constraints, fixed $k$ constraints, ``hedron" constraints, and mass gap constraints. The fixed $q$ constraint is given by the Hankel matrices $H(g_{k,q^*})$ and the shifted Hankel matrices being PSD for arbitrary fixed $q^*$. This is the necessary and sufficient condition that $g_{k,q^*}$ is a positive span of moment curve of fixed $q^*$:
\eq g_{k,q*}=\sum_a p_{a,q*}\frac{1}{m_a^{2k}}.\nn \eqe
The fixed $k$ constraint is the positivity of the polytopal wall $\langle\vec{g}_k,\vec{v}_i,\vec{v}_{i+1}...\rangle\ \geq0$. It is the necessary and sufficient condition that $\vec{g}_{k^*}$ lives in the convex hull of formed by the vectors defined by Gegenbauer coefficients, $\vec{v}_{l}$, for arbitrary fixed $k^*$:
\eq g_{k^*,q}=\sum_a p_{a,k^*} v_{l_a,q}. \nn \eqe
The hedron constraint is that the Hankel matrices $H(\langle\vec{g_k},\vec{v_l},\vec{v_{l+1}}...\rangle)$ and the shifted Hankel matrices being PSD. For example, the hedron constraint for a three-dimensional $\vec{g}_k$ would be
\eq \begin{pmatrix}
\langle\vec{g}_k,\vec{v}_l,\vec{v}_{l+1}\rangle & \langle\vec{g}_{k+1},\vec{v}_l,\vec{v}_{l+1}...\rangle & \hdots\\
\langle\vec{g}_{k+1},\vec{v}_l,\vec{v}_{l+1}\rangle & \langle\vec{g}_{k+2},\vec{v}_l,\vec{v}_{l+1}...\rangle & \hdots\\
\vdots & \vdots & \ddots
\end{pmatrix}\geq0. \eqe
These are necessary conditions for $g_{k,q}$ to live in the product space of the convex hull of moment curves and the convex hull of Gegenbauer-coefficient vectors. Finally, the mass gap condition is the positivity of the twisted Hankel matrices and the shifted ones mentioned in the previous sections being PSD.

We compare the $(\tilde{g}_{3,1}, \tilde{g}_{3,0})$ space carved out by the $k=2 \sim 5$ constraints in \cite{Arkani-Hamed:2020blm} with the one carved out by the new bounds derived the previous section, and with SDPB. Note that at $k=5$, the hedron constraints are trivial. The first non-trivial one emerges at $k = 6$, and it is computationally prohibitive to analyze fully. The $k=2 \sim 5$ constraints of the s-channel geometry in the previous paper are listed as follows.
\begin{enumerate}
	\item Fixed $q$:
	\eq \label{fixed-q}\begin{pmatrix}
			g_{2,0} & g_{3,0}\\
			g_{3,0} & g_{4,0}\\
		\end{pmatrix},
		\begin{pmatrix}
			g_{3,0} & g_{4,0}\\
			g_{4,0} & g_{5,0}\\
		\end{pmatrix},
		\begin{pmatrix}
			g_{3,1} & g_{4,1}\\
			g_{4,1} & g_{5,1}\\
		\end{pmatrix},\ g_{4,1},\ g_{4,2},\ g_{5,2},\ g_{5,3}\ \geq 0 \eqe
	\item Fixed $k$: 
	\eq \label{fixed-k} \langle\vec{g}_4,\vec{v}_{l},\vec{v}_{l+1}\rangle,\ \langle\vec{0}, \vec{g}_5 ,\vec{v}_{l}, \vec{v}_{l+1}\rangle,\ \langle\vec{g}_5 ,\vec{v}_{l}, \vec{v}_{l+1},\vec{\infty}\rangle\ \geq 0
	\eqe
	\item Mass gap:
	\eqa \label{mass-gap}
		\left(g_{4,2} - g_{5,2}\right), \left(g_{3,0} - g_{4,0}\right),\left(g_{3,1} - g_{4,1}\right),\left(g_{4,1} - g_{5,1}\right) \geq 0 \nn \\ 
		\begin{pmatrix}
			g_{2,0} - g_{3,0} & g_{3,0} - g_{4,0}\\
			g_{3,0} - g_{4,0} & g_{4,0} - g_{5,0}\\
		\end{pmatrix} \geq 0 \nn \\
		\left(g_{2,0}-2g_{3,0}+g_{4,0}\right), \left(g_{3,0}-2g_{4,0}+g_{5,0}\right) , \left(g_{3,1}-2g_{4,1}+g_{5,1}\right) \geq 0 \nn\\
	\left(g_{2,0}-3g_{3,0}+3g_{4,0}-g_{5,0}\right)\geq 0 \eqae
\end{enumerate}
With these constraints and the $k=5$ null constraint $g_{5,2}=g_{5,3}$, the allowed region of $(\tilde{g}_{3,1},\tilde{g}_{3,0})$ space can be carved out and the region is an infinite stripe \[0 \leq \tilde{g}_{3,1},\ 0 \leq \tilde{g}_{3,0} \leq 1.\] 
The fact that \(g_{3,1}\) has no upper bound can be easily seen from the constraints (\ref{fixed-q}), (\ref{fixed-k}), and (\ref{mass-gap}). In this paper, the upper bound of $g_{3,1}$ arises from the bi-variate moment matrix eq.(\ref{s moment}) being PSD, leading to a finite region. In \cite{Arkani-Hamed:2020blm}, the Gegenbauer coefficients were not identified as a GL transform of a moment curve and therefore the previous constraints do not capture the geometry of the product moment curve, thus the constraints in the previous paper are necessary but not sufficient.
\begin{figure}[h]
	\centering
	\includegraphics[width=0.85\textwidth]{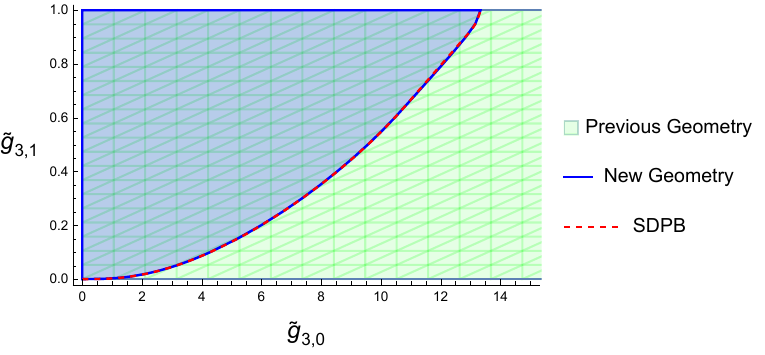}
	\caption{Comparison of geometry constraints in \cite{Arkani-Hamed:2020blm} and new geometry constraints up to $k=5$, and SDPB for $(\tilde{g}_{3,1},\tilde{g}_{3,0})$. All plots are using $k=5$ null constraint due to cyclic invariance.}
\end{figure}

\section{The EFThedron}\label{sec4}

We now turn to the more general set up, the case where both $s$ and $u$-channel singularities are allowed at fixed $t$. This set up will be applicable to general $a,b\rightarrow a,b$ scattering where $a,b$ can be distinct external states. If we consider the explicit amplitude $M(a, b,b,a)$, such that the $t\rightarrow0$ limit corresponds to forward scattering, the low energy couplings, defined as 
\eq
M^{\rm IR}(a, b,b,a)=\sum_{k,q}\;g_{k,q}z^{k{-}q} t^q\,,
\eqe
where $s={-}t/2{+}z$, $u={-}t/2{-}z$. Due to the $2\leftrightarrow 3 $ symmetry of the setup, the couplings will be non-zero only for even powers of $z$, i.e. $k{-}q\in even$. The dispersive representation equates
\eq
\sum_{k,q}\;g_{k,q}z^{k{-}q} t^q=-\sum_i p_i \, P_{\ell_i}(1{+}\frac{2t}{m^2_i})\left(\frac{1}{-\frac{t}{2}-z-m_i^2}{+}\frac{1}{-\frac{t}{2}+z-m_i^2}\right)\,,
\eqe
where once again, the equality is understood in terms of a Taylor expansion in $z,t$ on both sides and for terms with $k{-}q\geq2$. The result is the identity
\eq\label{uGeo}
 g_{k,q} = \sum_{i} p_i \frac{u_{\ell_i, k,q}}{m^{2(k{+}1)}_i}
\eqe
where $u_{\ell, k,q}$ are related to the Legendre coefficients $v_{\ell,q}$ via an $k$-dependent degenerate GL rotation ${\bf{G}}^k_{u,v}$ 
\eq\label{uvec}
\vec{u}_{\ell,k}=\left(\begin{array}{c}u_{\ell,k,0} \\0 \\u_{\ell,k,2}\\ \vdots \\0\\ u_{\ell,k,k}\end{array}\right)
	=
	\left(\begin{array}{cccccc}
	1 & 0 & 0 & 0 & 0 & 0 \\0 & 0 & 0 & 0 & 0 & 0 \\ 
	\frac{(k{-}1)_{2}}{2}\frac{1}{2^2}& -\frac{(k{-}1)_{1}}{2} & 1 & 0 & 0 & 0 \\ 
	\vdots & \vdots & \vdots & \vdots & \vdots & \vdots \\
	0 & 0 & 0 & 0 & 0 & 0 \\
	\frac{(1)_k}{k!}2^{-k} & -\frac{(1)_{k{-}1}}{k{-}1!}2^{1-k} & \frac{(1)_{k{-}2}}{k{-}2!}2^{2-k} & \cdots & -\frac{1}{2} & 1\end{array}\right)\left(\begin{array}{c}v_{\ell,0} \\ v_{\ell,1} \\v_{\ell,2}\\ \vdots \\v_{\ell,k{-}1}\\ v_{\ell,k}\end{array}\right)\,.
\eqe 
In the following, we will demonstrate how via a series of GL rotations, the above geometry can be converted to our familiar $a$-geometry.

\subsection{The EFThedron as a projection}
Due to the $k$-dependence of ${\bf{G}}^k_{u,v}$, the convex hull is now of an entangled product geometry. In~\cite{Arkani-Hamed:2020blm}, the boundaries of this space were obtained by approximating the geometry with a series of deformed moment curves.

Before the GL transformation, the hull is our previous $a$-geometry. Thus a priori, simply acting with an inverse transform on $g_{k,q}$ in eq.(\ref{uGeo}), we can revert to the $a$-geometry and apply the associated Hankel positivity. The obstruction is of course that ${\bf{G}}^k_{u,v}$ is degenerate, of half-rank. To this end, lets instead consider an auxiliary $\widetilde{\bf{G}}^k_{u,v}$ 
 \eq\label{uvec}
\widetilde{\bf{G}}^k_{u,v}=\begin{pmatrix}1 & 0 & 0 & 0 & &0 & &0 \\0 &{ \color{red}\textbf{1}} & 0 & 0 && 0 && 0 \\ \frac{(k{-}1)_{2}}{2}\frac{1}{2^2}& -\frac{(k{-}1)_{1}}{2} & 1&  0 & &0 && 0\\0 & 0 & 0 & { \color{red}\textbf{1}}  && 0 && 0 \\ \vdots & \vdots & \vdots & \vdots && \vdots && \vdots \\0 & 0 & 0 & 0 && { \color{red}\textbf{1}}  && 0 \\\frac{(1)_k}{k!}2^{-k} & -\frac{(1)_{k{-}1}}{k{-}1!}2^{1-k} & \frac{(1)_{k{-}2}}{k{-}2!}2^{2-k} & \cdots && -\frac{1}{2} && 1\end{pmatrix}\,.
\eqe 
where we have inserted auxiliary rows such that $\widetilde{\bf{G}}^k_{u,v}$ becomes full rank. Defining 
\eq
\vec{u}'_{\ell,k}=\widetilde{\bf{G}}^k_{u,v}\vec{v}_{\ell}\,,
\eqe
we see that $\vec{u}'_{\ell,k}=\vec{u}_{\ell,k}$ for the components where $k{-}q=even$, while $\vec{u}'$ will have non-zero entries for $k{-}q=odd$. Now let's consider 
\eq
 \vec{g}'_{k} = \sum_{i} p_i \frac{\vec{u}'_{\ell_i, k}}{m^{2(k{+}1)}_i}=\sum_{i} p_i \frac{1}{m^{2(k{+}1)}_i}\widetilde{\bf{G}}^k_{u,v}\vec{v}_{\ell_i}\,.,
\eqe
where $\vec{g}'_{k}$ now contains $k{-}q=odd$ auxiliary couplings. Multiplying both sides by $(\widetilde{\bf{G}}^k_{u,v})^{{-}1}$ and ${\bf{G}}_{J,v}$, we can now recover the $a$-geometry
\eq
\left(\begin{array}{cccc}a_{2,0} & \; & \; & \; \\ a_{3,0} & a_{3,1} & \; & \; \\ a_{4,0} & a_{4,1} & a_{4,2} & \; \\ \vdots & \vdots & \vdots &  \vdots \end{array}\right)={\bf{G}}_{J,v}(\widetilde{\bf{G}}^k_{u,v})^{{-}1} \vec{g}'_{k}=\sum_i \frac{p_i}{m^6_i} \left(\begin{array}{c} 1 \\  \frac{1}{m^2_i} \\  \frac{1}{m^4_i} \\  \vdots\end{array}\right)\otimes (1, J^2_i, J_i^4,\cdots)\,.
\eqe
In other words, \textit{the full EFThedron is simply a projection of the $a$-geometry}! 

In practice, we first take the Hankel positivity constraints on the couplings in the $a$-geometry, $a_{k,q}$, and GL rotate back to $g'_{k,q}$ through the action of $\widetilde{\bf{G}}^k_{u,v}( {\bf{G}}_{J,v})^{{-}1}$. Second, project out the auxiliary couplings from the constraint, ie. those which have $k{-}q=odd$. Importantly, the result is independent of how we complete the rank for ${\bf{G}}^k_{u,v}$, i.e. after projecting out the auxiliary couplings we retain a ``gauge" invariant result. It is straight forward to verify that this is indeed the case.

For $a,b\rightarrow b,a$ scattering, the constraint imposed by the $a$-geometry is the final result. However, for identical scalars ($a=b$), the low energy amplitude must now respect permutation invariance, i.e. the EFT amplitude is a function of $\sigma_2=s^2+t^2+u^2$ and $\sigma_3=s t u$, implying linear relations between $g_{k,q}$ of equal $k$, i.e. the couplings live on the permutation plane
\eq
M(z,t)=M(\frac{3t}{4}{+}\frac{z}{2},z{-}\frac{t}{2}),\quad 
\eqe
Thus for identical scalar scattering, the geometry we are interested in is the intersection of the projected $a$-geometry, intersected with the permutation plane.
\subsection{Explicit region analysis}
The conditions we will use are mostly identical to those for the $s$-EFThedron. The two differences are related to polytopes, as now $J^2=\{0,6,20,\ldots\}$, and the null constraint, which now first appears at $k=4$. Because the summand of the null constraint, given by
\eq
n_4=\sum_i \frac{p_i}{m_i^{10}} J^2_i(J_i^2-8)
\eqe
has a root at $J^2=8$, only the polytope wall $\langle \vec{a}_4,6,20\rangle$ will intersect the symmetry plane. 

Like for the $s$-EFThedron,  we wish to bound $(a_{4,0},a_{3,1})$ space. Because the null plane is at order $k=4$, we  we will be able to analytically solve the constraints. First, let us impose all possible constraints up to order $k=4$
\eqa
&& a_{2,0}-a_{3,0},a_{3,0}-a_{4,0},a_{3,1}-a_{4,1}\ge0\label{easy2}\\
&&\begin{pmatrix} 
    a_{2,0} & a_{3,0}&a_{3,1} \\
    a_{3,0} & a_{4,0}&a_{4,1}\\
    a_{3,1}& a_{4,1} &a_{4,2}
    \end{pmatrix}\ge 0\\
&&     \textrm{det} \begin{pmatrix}
    a_{4,0} & 1 & 1\\
    a_{4,1} & 6 & 20\\
    a_{4,2} & 36 & 400
    \end{pmatrix}\ge 0\label{polytope4}
\eqae
subjected to the null constraint $a_{4,2}=8a_{4,1}$. We find 
    \begin{align}
  \nonumber  &0<{\tilde{a}_{3,1}}< \sqrt{\frac{160}{3}{\tilde{a}_{4,0}}} \quad &\textrm{for}&\quad   0<{\tilde{a}_{4,0}}\leq \frac{5}{6}, \\
&   0<{\tilde{a}_{3,1}}<\sqrt{\frac{80}{9}\left({\tilde{a}_{4,0}}-{\tilde{a}_{4,0}}^2\right)}+\frac{20}{3} \tilde{a}_{4,0}\quad   &\textrm{for}&\quad   \frac{5}{6}<{\tilde{a}_{4,0}}<1
   \label{discri}
    \end{align}    
    
We can further impose the positivity of the following matrices involving $k=5$ couplings:
\eq
\left(\begin{array}{ccc}a_{3,0} & a_{4,0} & a_{4,1} \\ a_{4,0} & a_{5,0} & a_{5,1} \\ a_{4,1} & a_{5,1} & a_{5,2}\end{array}\right),\left(\begin{array}{ccc}a_{3,1} & a_{4,1} & a_{4,2} \\ a_{4,1} & a_{5,1} & a_{5,2} \\ a_{4,2} & a_{5,2} & a_{5,3}\end{array}\right),\;\left(\begin{array}{lll}a_{2,0}-a_{3,0} & a_{3,0}-a_{4,0} & a_{3,1}-a_{4,1} \\ a_{3,0}-a_{4,0} & a_{4,0}-a_{5,0} & a_{4,1}-a_{5,1} \\ a_{3,1}-a_{4,1} & a_{4,1}-a_{5,1} & a_{4,2}-a_{5,2}\end{array}\right)
\eqe
as well as the polytopal constraints,
\eq
\langle 0,a_5,6,20\rangle,\quad\langle a_5, 6, 20,\infty\rangle\,.
\eqe
Solving these conditions, we obtain the final region is given by
\begin{align} \label{complete}
  \nonumber & 0<{\tilde{a}_{3,1}}< \sqrt{\frac{160}{3}{\tilde{a}_{4,0}}} & \textrm{for}&\quad   0<{\tilde{a}_{4,0}}\leq \frac{27}{40}, \\
 &  0<{\tilde{a}_{3,1}}< \frac{20}{21} \left(6 \tilde{a}_{4,0} + \sqrt{(21 - 20 \tilde{a}_{4,0}) \tilde{a}_{4,0}}\right)   &\textrm{for}&\quad   \frac{27}{40}<{\tilde{a}_{4,0}}<1
 \end{align}
which is stronger than using only the $k\le 4$ constraints, and we find it matches exactly to numerical results.

\paragraph{Real geometry}
The last step requires us to rotate from the $a$ to the $g$ couplings, and project out couplings $g_{k,q}$ with $k+q=\textrm{odd}$. In our low dimension example the only non-trivial rotation is $a_{3,1}\rightarrow \frac{3}{2}g_{3,0}+g_{3,1}$. Doing these operations we can explicitly solve the conditions up to $k=5$ described above, and we obtain the complete $(g_{3,1},g_{4,0}$) allowed region,  
\begin{align}
\textrm{Region I:}\quad \tilde{g}_{31} &=-\frac{3}{2} \sqrt{\tilde{g}_{40}},  &&0 \leq \tilde{g}_{40} \leq 1 \\ 
\textrm{Region II:}\quad \tilde{g}_{31} &=\frac{1}{2} \sqrt{\frac{427}{3} \tilde{g}_{40}}, &&0 \leq \tilde{g}_{40} \leq \frac{243}{427}\\ 
\textrm{Region III:}\quad \tilde{g}_{31} &=\frac{30}{7} \tilde{g}_{40}+\frac{37}{42} \sqrt{\tilde{g}_{40}\left(21-20 \tilde{g}_{40}\right)},  &&\frac{243}{427} \leq \tilde{g}_{40} \leq 1
\end{align}
\begin{figure}[h] 
 \centering
  \includegraphics[width=0.95\textwidth]{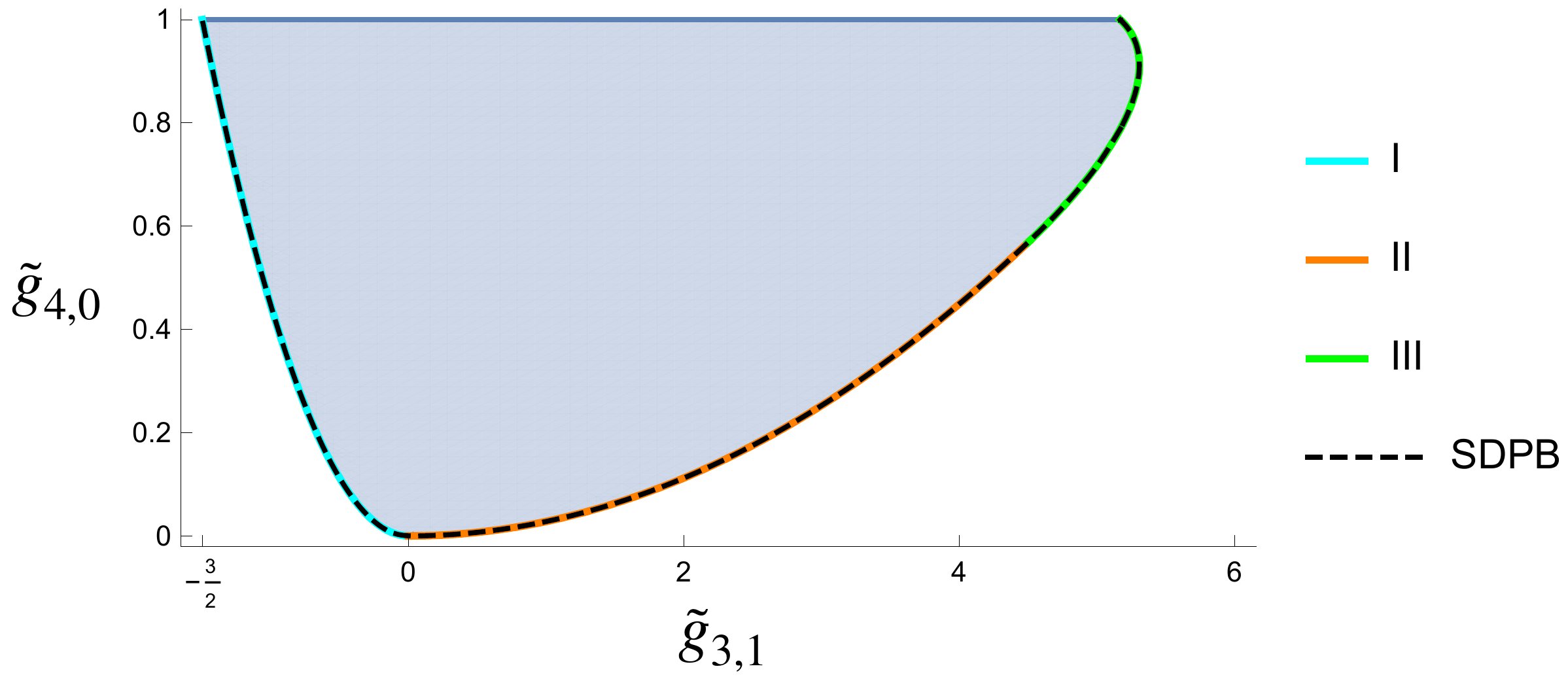} 
   \caption{($\tilde{g}_{3,1}$, $\tilde{g}_{4,0}$) space with $k=4$ null constraint, geometry vs SDPB}
 \label{g31g400}
\end{figure}
This is shown in Figure \ref{g31g400}, and we find it matches perfectly the results from SDPB~\cite{Caron-Huot:2020cmc}. For a more precise comparison, we can compute the upper bound on $\tilde{g}_{3,1}$, which we find to be
\eq
2\tilde{g}_{3,1}<\frac{9}{2} + \frac{7}{4} \sqrt{\frac{61}{5}}\approx 10.6125
\eqe 
identical to the value found with SDPB.

The boundary structure is the following
\begin{itemize}
\item Region I
\eq
a_{3,1}=0, \quad \textrm{Rank}\left(\begin{array}{cc}a_{2,0} & a_{3,0}  \\ a_{3,0} & a_{4,0} \end{array}\right)=1
\eqe

\item Region II
\eq
 \langle \vec{a}_4,6,20\rangle=0,\quad  \textrm{Rank}\left(\begin{array}{cc}a_{2,0} & a_{3,1}  \\ a_{3,1} & a_{4,2} \end{array}\right)=1
\eqe

\item Region III
\begin{align}\nonumber
&\langle \vec{a}_4,6,20\rangle=0,\quad \langle 0,\vec{a}_5,6,20\rangle=0,\quad\langle \vec{a}_5, 6, 20,\infty\rangle=0\\
\nonumber &\textrm{Rank}\left(\begin{array}{ccc}a_{3,0} & a_{4,0} & a_{4,1} \\ a_{4,0} & a_{5,0} & a_{5,1} \\ a_{4,1} & a_{5,1} & a_{5,2}\end{array}\right)=2,\quad \textrm{Rank}\left(\begin{array}{ccc}a_{3,1} & a_{4,1} & a_{4,2} \\ a_{4,1} & a_{5,1} & a_{5,2} \\ a_{4,2} & a_{5,2} & a_{5,3}\end{array}\right)=2,\\
&\textrm{Rank}\left(\begin{array}{lll}a_{2,0}-a_{3,0} & a_{3,0}-a_{4,0} & a_{3,1}-a_{4,1} \\ a_{3,0}-a_{4,0} & a_{4,0}-a_{5,0} & a_{4,1}-a_{5,1} \\ a_{3,1}-a_{4,1} & a_{4,1}-a_{5,1} & a_{4,2}-a_{5,2}\end{array}\right)=1
\end{align}
\end{itemize}


\section{Constraints on the UV spectrum}\label{UV}
\begin{figure}[h] 
   \centering
   \includegraphics[width=3.3in]{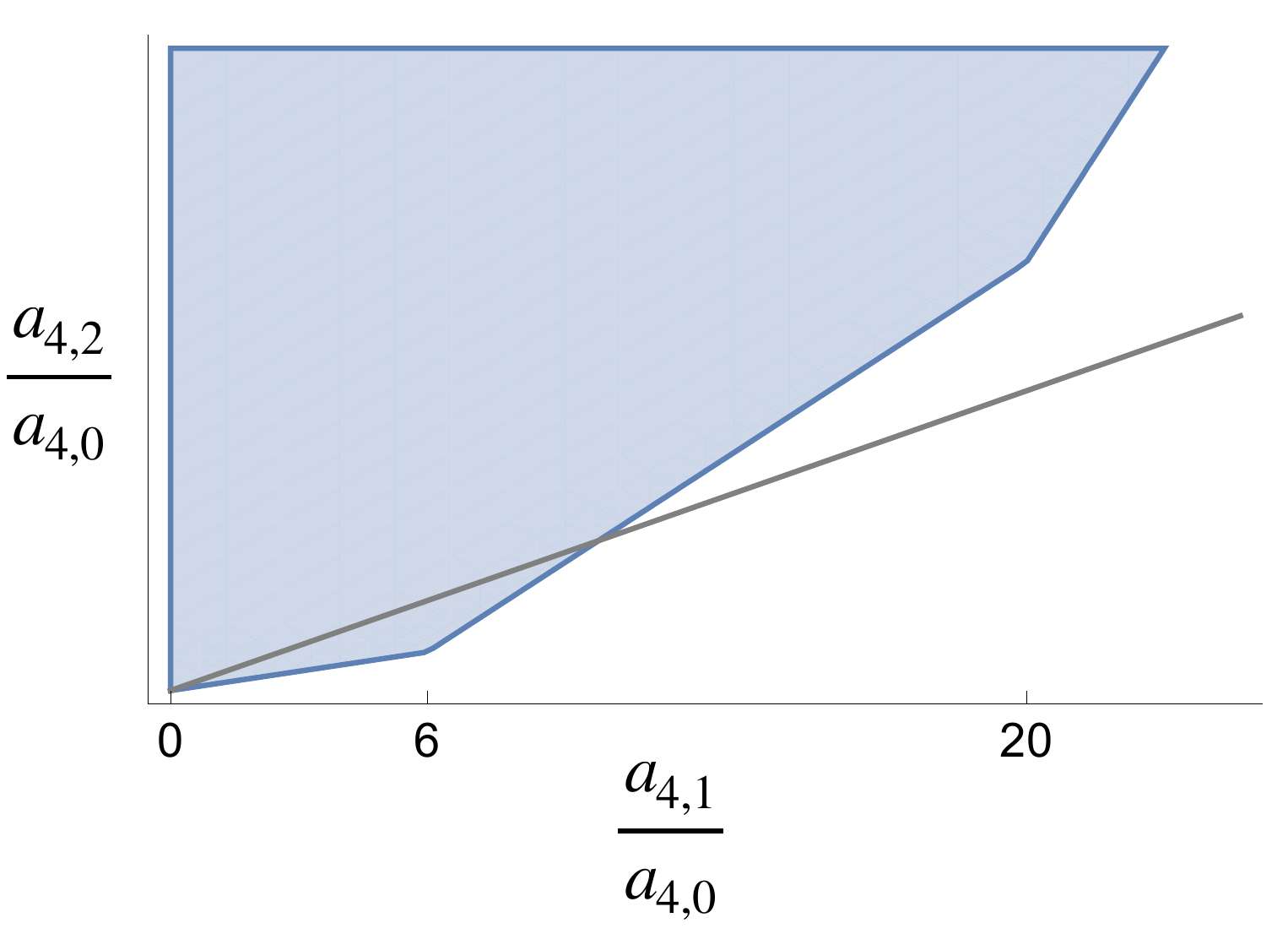} 
   \caption{$(a_{4,1},a_{4,2})$ polytope and $a_{4,2}=8a_{4,1}$ symmetry plane}
   \label{2dplotb}
\end{figure}

As we have seen in previous analysis, while the space for the hull is in general unbounded its intersection with symmetry planes yields a finite region.  The reason that it is finite is simple, in general only the low spins are one side of the symmetry plane while the remaining majority are on the other side. Indeed take for example the $su$-channel $k=4$ polytope intersected with the symmetry plane $a_{4,2}=8a_{4,1}$, which we plot again in Figure \ref{2dplotb}.
We see that only the spin 2 ($J^2=6$) vertex lives on one side of the plane. This immediately tells us that the $k=4$ null constraint requires spin 2 to be present. In other words, to live on the symmetry plane, certain distributions in the spin spectrum are necessary.

A convenient way of formulating such conditions is the statement that the hull must have zero components perpendicular to the symmetry plane. This condition is often termed ``null constraints", which we denote as $n_k$. The subscript $k$ indicates that it associated with symmetry constraints on the $k$-th level couplings $g_{k,q}$. 
Once we substitute the dispersive representation for the couplings, the condition translates into the vanishing of a sum over the UV states. For example,  the symmetry plane in Figure  \ref{2dplotb} corresponds to a null condition
\eq
n_4=a_{4,2}-8a_{4,1}=\sum_i \frac{p_i}{m_i^{10}}\left(J_i^4-8J^2_i\right)
\eqe
In general, null conditions will have a form
\eq\label{omega}
n_k=\sum_i p_{i}\frac{1}{(m_i^2)^{k{+}1}}\omega_k(\ell_i)=0
\eqe   
where $\omega_k(\ell)$ is a polynomial in $\ell$, and there may be several null constraints at the same $k$ level. The fact that the above expression must sum to zero imposes non-trivial conditions on the UV spectrum. We will analyze these constraints in steps, first utilizing just the sign of $\omega_k(\ell)$, then its magnitude, and finally linear combinations of $\omega_k(\ell)$ with distinct $k$. We summarize our finding in the following

\begin{itemize}
  \item \textbf{Conditions on spins} Given that $\omega_k(\ell)$ is a finite degree polynomial, at sufficiently large spins its sign is fixed, which we can always take to be positive. Thus if the spectrum only contains high spin states for which $\omega_k(\ell)>0$, it is impossible for the null constraints to be satisfied, since negative terms must also be present. This tells us that the real roots of $\omega_k(\ell)$ encode the necessary conditions the spin distribution must satisfy for $n_k$ to hold. Collecting the conditions from different $k$ one obtains global constraints on the spectrum. For example, we will find that assuming the existence of non-scalar states, \textit{the UV completion of a theory containing identical scalars must include an infinite number of arbitrarily high even spins, including all even spins $\ell\le  28$}. Certainly such result is not surprising and can be argued directly from the Froissart-Gribov formula, i.e. it is inconsistent to have higher spins without having the complete even spin spectrum.\footnote{We thank the referee for pointing this out.} The exercise here is to see how such conditions arrises from geometry, and set stage to the following.  
  
  \item  \textbf{Conditions on probabilities} Beyond the sign pattern, the functional form $\omega_k(\ell)$ tells us the magnitude of the contribution for each spin. Since the contribution must cancel out, this imposes upper bounds on the averaged spinning spectral function $\langle p_{\ell,k}\rangle$ for any fixed $\ell$. In particular, for identical scalars we can deduce upper bounds for the ratio $\langle p_{\ell,k}\rangle/\langle p_{2,k}\rangle$, such as
  \eq
  \frac{\langle p_{4,\ell}\rangle}{\langle p_{4,2}\rangle}\leq\frac{12}{\ell (\ell+1) \left(\ell^2+\ell-8\right)},~(\ell\geq4)
  \eqe
  
  \item \textbf{Conditions on masses} Combining different $k$-order constraints, we now obtain functions that are no longer homogenous in $m_i$. This will then allow us to deduce that there must exists a state in the spectrum whose mass is below a ``gap" 
  \eq
  m_\ell^2\leq M^2 B(\ell)
  \eqe
  where $M$ is the heaviest spin 2 particle.
    \end{itemize}

In the following we will give the details for the $s$- and full EFThedron, as well as the case when the external states carry spin.

\subsection{Implications for spin spectrum}\label{sec:spin spectrum}Substituting the dispersive representation for the couplings in the null constraints, we will in general arrive at
\eq\label{null}
n_k=\sum_{i} \frac{p_{\ell_i}}{(m_i^2)^{k{+}1}}\omega_k(\ell_i)=0\,,
\eqe
where the function $\omega_k(\ell)$ is a polynomial in $\ell$ and thus have finite roots. Since $p_\ell$ is positive, this immediately tells us there must be states residing in both the positive and negative regions of the polynomial. We begin by first analyzing this sign pattern for the $s$-EFThedron.

\noindent {\bf $s$-EFThedron}

Imposing cyclic symmetry
\begin{align}\label{cyc}
M(s,t){=}M(t,s)\quad \rightarrow\quad g_{k,q}=g_{k,k{-}q}\,,
\end{align}
leads to the following null constraints
\begin{align}
g_{k,q}{-}g_{k,k-q}=\sum_i \frac{p_i}{(m^2_i)^{k{+}1}}\left(v_{\ell_i,q}-v_{\ell_i,k-q} \right)=0\,.
\end{align}
Since the couplings are well defined only for $k\ge 2$ and $k-q\ge 2$, the first valid cyclic null constraint appears at $k=5$, and is given by
\begin{align}
n_5=g_{5,2}-g_{5,3}=\sum_i \frac{p_i}{m_i^{12}}(\ell_i{-}1) \ell_i (\ell_i+1) ( \ell_i+2) (\ell_i^2 {+}\ell_i{-}15 ) =0\,.
\end{align}
First, if all spins in the theory are 0 or 1, the null constraint is trivially satisfied. However, if there is at least one particle with spin $\ell>1$, the null constraint becomes non-trivial. Importantly we must have states residing in both positive and negative regions of the polynomial $w_5(\ell)=(\ell{-}1) \ell (\ell+1) (\ell+2) ( \ell^2{+}\ell-15)$, in order for the null constraint to hold. Noting the last factor has a positive root at $\ell\sim3.4$, lets consider the sign pattern of this function for each spin 
\begin{center}
\begin{tabular}{ |c|c|c|c|c|c|c|c| } 
 \hline
$\ell$& 0 & 1&2&3&4&5&$\ldots$ \\ 
 \hline
$\omega_5(\ell)$ & 0 & 0&$-$&$-$&$+$&+&+ \\ 
 \hline
\end{tabular}
\end{center}
 We observe that for spins 2 and 3 it is negative, while for spins 4 and greater it is  positive. This means for a theory to satisfy cyclic invariance and contain spins beyond spin 1, it must contain either a spin 2 or a spin 3, as well as at least one spin 4 or higher.

The $k=6$ null constraint is weaker and does not add new information, while at $k=7$ the null constraint $g_{7,3}=g_{7,4}$ yields
      \eq
\omega_7(\ell)=    ( \ell-2) ( \ell-1) \ell ( \ell+1) ( \ell+2) ( \ell+3) (  \ell^2+\ell-28)\, ,
      \eqe
with a final root at  $\ell \sim 4.8$, and the following sign pattern
         \begin{center}
\begin{tabular}{ |c|c|c|c|c|c|c|c| } 
 \hline
$\ell$& 0 & 1&2&3&4&5&$\ldots$ \\ 
 \hline
$\omega_7(\ell)$ & 0 & 0&0&$-$&$-$&+&+ \\ 
 \hline
\end{tabular}
\end{center}
This tells us that if we have states with $\ell\geq3$, then the spectrum must have either a spin 3 or 4 and at least one spin $\ge 5$.

     We can continue this argument order by order at higher $k$. It is useful to specifically consider constraints $g_{2a{+}1,a}=g_{2a{+}1,a{+}1}$ which takes the form
      \begin{align}
      \omega_{2a+1}(\ell)=(\ell^2+\ell-f_c(a)) \prod_{i=0}^{a-1} (\ell-i) \prod_{j=1}^{a+1} (\ell+j)
      \end{align}
      where $f_c(a)=(a+1)(2a+1)$, leading to a positive root at
      \eq  
  \ell_c(a) \equiv \frac{  {-}1{+}\sqrt{1{+}4 f_c(a)}}{2}
      \eqe
      which for integer $a>0$ is well approximated by 
      \eq
      \ell_c(a)\approx  \sqrt{2} a +\frac{1}{2} \left(\frac{3}{ \sqrt{2}}-1 \right)
      \eqe
      
Now  $\omega_{2a+1}(\ell)$ has zeroes for $\ell\leq a{-}1$, and has only one negative region between $a-1 < \ell<  \ell_c(a)$ , and becomes positive for $\ell >   \ell_c(a)$.  This leads to the following conditions
\begin{align}
k&=5:\ell\in \{2,3\} \textrm{ and } \ell\in \{4,5,\dots\}\nonumber\\
k&=7:\ell\in \{3,4\}  \textrm{ and } \ell\in \{5,6,\dots\}\nonumber\\
k&=9:\ell\in \{4,5,6\}  \textrm{ and }  \ell\in \{7,8,\dots\}\nonumber\\
k&=11:\ell\in \{5,6,7\}  \textrm{ and } \ell\in \{8,9,\dots\}\,.
\end{align}
These constraints are valid for UV completions of colored scalars states involving massive states beyond spin 1. From the above we can immediately conclude that the spectrum must be unbounded in spin. As $k$ increases, we observe that the conditions on spins become weaker. This is because the interval between the last two roots grows as $\sim (\sqrt{2}-1)a$. The same behavior will also occur for the $su$-EFThedron, but the conditions will be stronger.

\noindent {\bf The full EFThedron}

We now perform the same analysis with permutation invariance of identical scalars. Recall that the permutation plane for the full EFThedron is defined as
\begin{align}\label{per}
M(z,t)=M(\frac{z}{2}{+}\frac{3t}{4},{-}\frac{t}{2}{+}z)\,.
\end{align}
When translated into linear relations for the couplings, for the first few $k$s they are given by (recall that we must have $k\ge 2$, $k-q\ge 2$ )
\begin{align}\label{eqn:sunull}
k=4:\quad g_{4,2}&=\frac{2}{3}g_{4,0},\quad &k=5:&\quad g_{5,3}=\frac{g_{5,1}}{2}\nonumber\\
k=6:\quad g_{6,4}&=\frac{45}{16}g_{6,0}{-}\frac{1}{2}g_{6,2},\quad &k=7:&\quad g_{7,3}=\frac{4}{5}g_{7,1},\quad  g_{7,5}=\frac{16}{3}g_{7,1}\quad\nonumber\\
k=8:\quad g_{8,4}&=\frac{21}{8}g_{8,0}{-}\frac{1}{4}g_{8,2},&&\quad g_{8,6}=\frac{21}{8}g_{8,0}{-}\frac{5}{16}g_{8,2}\,.
\end{align}

Substituting the dispersive representation for the $k=4$ the null constraint one finds
\begin{align}
n_{4}=\sum_i \frac{p_i}{m^{10}}\ell_i(\ell_i+1)(\ell_i^2 +\ell_i-8)=0\,.
\end{align}
The sign pattern of the summand $\omega_4(\ell)=(\ell(\ell+1)-8)$ is given as
          \begin{center}
\begin{tabular}{ |c|c|c|c|c|c| } 
 \hline
$\ell$& 0 &2&4&6&\ldots \\ 
 \hline
$\omega_4(\ell)$ & 0 & $-$&+&+&+ \\ 
 \hline
\end{tabular}
\end{center}
This implies that if we have states other than scalars, then we must have a spin $\ell=2$, and at least one state with $\ell\ge 4$. Going to higher $k$, our strategy will be to take linear combinations of null constraints that force roots for lower spins, are negative for one particular spin, and positive for higher spins. At $k=5,6$ no new constraints in terms of spectrum arise. At  $k=7$ we have two independent null constraints
\eq
n_{7}\equiv g_{7,3}-\frac{4}{5}g_{7,1}=0,\quad       n_{7}'\equiv g_{7,5}-\frac{16}{3} g_{7,1}=0
\eqe
Both have zeros at $l=0$, and we would like to have a root at $l=2$. The combination
\eq
n_7+n_7'=\sum_i \frac{p_i}{m_i^7} (\ell_i-2)\ell_i (\ell_i+1) (\ell_i+3) \left(\ell_i^2+\ell_i-\frac{49}{2}\right)=0
\eqe
achieves just that, and as desired we obtain sign pattern
           \begin{center}
\begin{tabular}{ |c|c|c|c|c|c| } 
 \hline
$\ell$& 0 &2&4&6&\ldots \\ 
 \hline
$\omega_7(\ell)$ & 0 & 0&-&+&+ \\ 
 \hline
\end{tabular}\,.
\end{center}
We observe it is negative only for $\ell=4$, and positive for $\ell\ge 6$. Thus now we have the new condition that the spectrum must contain at least one $\ell=4$ particle, as well even higher spin particles.

In general, it is easy to arrange null constraints at order $k=3a+1$ in the form
\eq
\omega_{3a+1}(\ell)=(\ell^2+\ell-f_p(a)) \prod_{i=0}^{a-1} (\ell-2i) \prod_{j=0}^{a-1} (\ell+2j+1)
\eqe
where $f_p(a)=( 3 a+1)^2/2$, leading to a root at 
\eq
\ell_p(a)\approx \frac{3 a}{\sqrt{2}}+\frac{1}{2} \left(\sqrt{2}-1\right)
\eqe 
Like in $s$-EFThedron, we have roots for $l\le 2a-2$, negative values for $2a-2< l<\ell_p(a)$, and positive values for $l> \ell_p(a)$. For constraints up to $k\le 43$,  this implies the following sign pattern
           \begin{center}
\begin{tabular}{ |c|c|c|c|c|c|c|c|c| } 
 \hline
$\ell$	&	0	&	2	&	\ldots	&	$2a-2$	&	$2a$	&	$2a+2$ &	\ldots \\ 
 \hline
$\omega_{3a+1}(\ell)$&	0	&	0	&	0		&	   0   	&	-	&	+	 &	+	\\ 
 \hline
\end{tabular}
\end{center}
which is negative only for $\ell=2a$, thus requiring the spin $\ell=2a$, as well as some state with $\ell\ge 2a+2$. Combining the different $k$s, we immediately conclude that \textit{for identical scalars coupled to at least a higher spin-state, the UV completion must contain an unbounded number of even spins, including all states below spin 28!} 

Going to $k=46$ we find that both spins 30 and 32 have negative contributions, so the presence of either is sufficient, a behavior that continues as we increase $k$. We see that permutation invariance imposes more stringent constraint on the EFThedron spectrum than cyclic invariance on the $s$-EFThedron, but the overall pattern is similar.

Note that pure scalar theories are not constrained, since scalars trivially solve the null constraints. This allows theories such as
\eq
M(s,t)=\frac{1}{s-m^2}{+}\frac{1}{t-m^2}{+}\frac{1}{u-m^2}\,
\eqe
to be compatible with permutation invariance while not containing any higher spin.

\subsection{Bounds on $\langle p_{k,\ell}\rangle$}
The function $\omega(\ell)$ not only puts constraints on the spin-spectrum, it also puts upper limits on how large the $p_\ell$s can take. It will be convenient to consider the \textit{average} spectral function $\langle p_{k,\ell}\rangle$, defined as 
\eq
\langle p_{k,\ell}\rangle\equiv\sum_{\{i,\ell_i=\ell\}}\frac{p_{\ell_i}}{m^{2(k{+1})}_i}\,.
\eqe
That is, we sum over contributions of all states with spin-$\ell$, weighted by their mass. In terms of $\langle p_{k,\ell}\rangle$, the dispersive representation take on a simple form
\eq
g_{k,q}=\sum_{\ell} \;\langle p_{k,\ell}\rangle u_{\ell,q}\,.
\eqe
To illustrate how constraints on $\langle p_{k,\ell}\rangle$ can be derived from symmetry considerations, starting with eq.(\ref{null}) we write the following sum rule
\eq
n_k=\sum_{i} \frac{p_{\ell_i}}{(m_i^2)^{k{+}1}}\omega_k(\ell_i)=\sum_{\ell}\langle p_{k,\ell}\rangle\omega_k(\ell)=0\,,
\eqe
where we have used the fact that $\omega_k(\ell)$ only depends on the spin and not the mass.

From the previous discussion, we have seen that as we tend to higher $k$, we have more than one null constraints $\omega_k(\ell)$. Suppose that we can find a linear combination $\tilde{\omega}_k(\ell)$ of null constraints, such that for some fixed spin $\ell^*$
\eq
\tilde{\omega}_k(\ell^*)<0,\quad \tilde{\omega}_k(\ell)\geq0 \;\forall\;\ell\neq\ell^*
\eqe
Now if we can conclude that $\ell^*$ must be part of the spectrum, then from the sum rule
\eq
\sum_{\ell} \;\langle p_{k,\ell}\rangle\tilde{\omega}_k(\ell)=0\,,
\eqe
we immediately find that there is an upper bound for the \textbf{ratio} $\langle p_{k,\ell}\rangle/\langle p_{k,\ell^*}\rangle$, in order to cancel the negative contribution from $\tilde{\omega}_k(\ell^*)$. This leads to 
\eq
\frac{\langle p_{k,\ell}\rangle}{\langle p_{k,\ell^*}\rangle}\leq\frac{|\tilde\omega_k(\ell^*)|}{\tilde\omega_k(\ell)},\quad \forall \ell\geq \ell_*\,.
\eqe
Note that this functional approach is exactly what has been implemented in the conformal bootstrap analysis~\cite{Rattazzi:2008pe,Caracciolo:2009bx}, and in particular to the problem of maximization of OPE coefficients. In some cases, such a functional cannot be found, and the best one can do is having a subset of spins being negative, with the remaining being non-negative. Then one has instead
\eq
\frac{\langle p_{k,\ell}\rangle}{\sum_{\ell^*}\langle p_{k,\ell^*}\rangle|\tilde\omega_k(\ell^*)|}\leq\frac{1}{\tilde\omega_k(\ell)}\,,
\eqe
where the sum is over the subset of spins whose $\tilde\omega_k(\ell^*)$ is negative.

\noindent \textbf{The $s$-EFThedron}

For $s$-EFThedron, the first non-trivial null constraint starts at $k=5$, with
\begin{equation}
n_5=\sum_{\ell}\langle p_{5,\ell}\rangle\omega_{5}(\ell)=0\,.
\end{equation}
The function $\omega_{5}(\ell)$ is negative for $\ell=\{2,3\}$ and positive for $\ell\geq4$, as illustrated in the following graph
$$\includegraphics[scale=0.4]{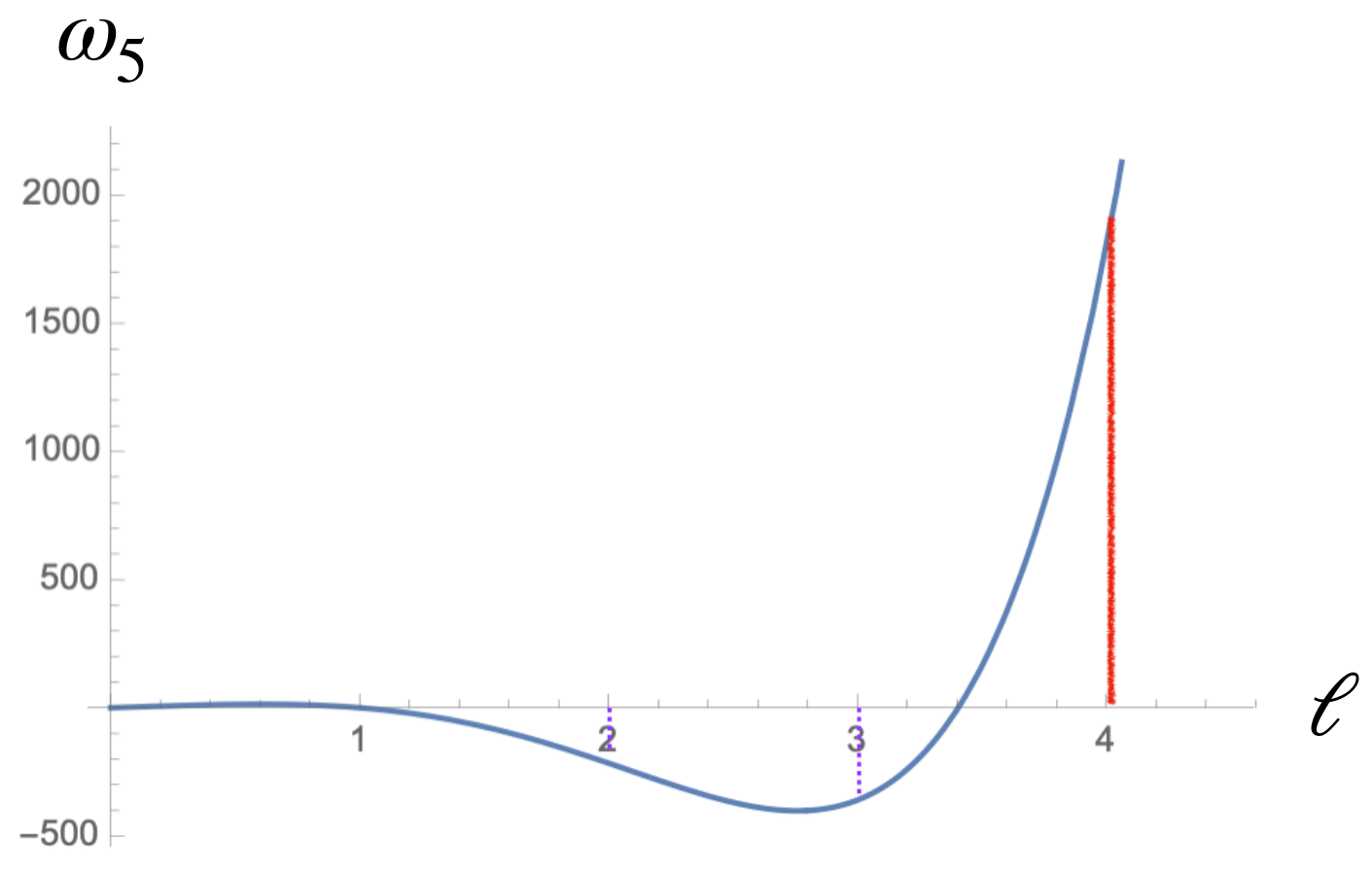}$$
The null constraint then translate to the sum of the positive region spin must equal to those in the negative region, weighted by their $\langle p_{5,\ell}\rangle$ respectively. For any given spin $\ell^*\geq 4$, the maximum value for $\langle p_{5,\ell^*}\rangle$ is attained when it is the sole source of positive contribution, which must match with the sum of that from spin-$2$ and $3$. Thus we have an upper bound of the form, 
\begin{equation}\label{eqn:seftk5}
\frac{\langle p_{5,\ell}\rangle}{\langle p_{5,2}\rangle|\tilde{\omega}_5(2)|{+}\langle p_{5,3}\rangle|\tilde{\omega}_5(3)|}{=}\frac{\langle p_{5,\ell}\rangle}{216\langle p_{5,2}\rangle{+}360\langle p_{5,3}\rangle}\leq\frac{1}{(\ell{-}1) \ell (\ell{+}1) (\ell{+}2)(\ell^2{+}\ell{-}15)},\;(\ell\geq4)\,.
\end{equation}

Moving to $k=6$, there is also only one null constraint in this case,
\begin{equation}
n_6=\sum_{\ell\in\mathbb{Z}^+}\langle p_{6,\ell}\rangle(\ell-1) \ell (\ell+1) (\ell+2) \left(\ell^4+2 \ell^3-17 \ell^2-18 \ell-72\right)=0
\end{equation}
The spin pattern is,
\begin{center}
\begin{tabular}{ |c|c|c|c|c|c|c|c|c| } 
 \hline
$\ell$	&	0	&	1	&	2	&	3	&	4	&	5 & 6 & 	\ldots \\ 
 \hline
$\tilde{\omega}_6(\ell)$&	0	&	0	&	- &	   -   	&	-	&	+	 &	+ & \ldots	\\ 
 \hline
\end{tabular}
\end{center}
Since the $k=5$ null constraint already requires either a spin 2 or spin 3 state, the new constraint does not impose anything new in terms of which spins should appear. However, it does yield a bound on $\langle p_{6,\ell}\rangle$:
\begin{equation}\label{eqn:seftk6}
\frac{\langle p_{6,\ell}\rangle}{3\langle p_{6,2}\rangle{+}15\langle p_{6,3}\rangle{+}10\langle p_{6,4}\rangle}\leq\frac{1152}{(\ell{-}1)\ell (\ell{+}1) (\ell{+}2) \left(\ell^4{+}2\ell^3{-}17\ell^2{-}18\ell{-}72\right)},\quad(\ell\geq5)\,.
\end{equation}

Starting from $k\geq7$, at each level of $k$ there will be more than one null constraint. Thus the general functionals $\tilde{\omega}_k(l)$ will take the linear combination of these null constraints. One can then obtain further bounds of the above form, where the spins of the positive region is bounded by the total contribution of the negative region.

\begin{figure}
\centering
\begin{minipage}[t]{0.78\textwidth}
\centering
\includegraphics[width=10cm]{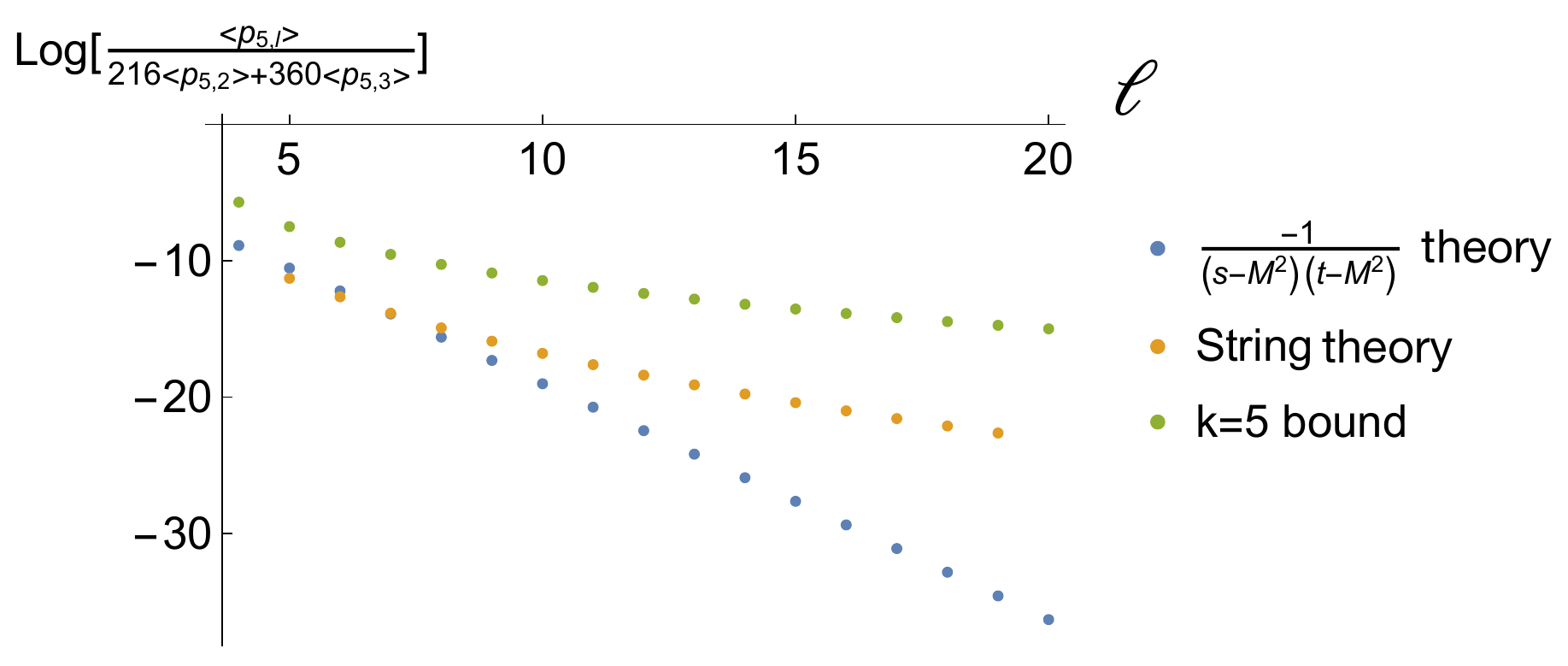}
\end{minipage}
\caption{Here we present the $k=5$ bounds in the $s$-EFThedron (\eqref{eqn:seftk5} \& \eqref{eqn:seftk6}) with the $S$-matrix $\frac{-1}{(s-M^2)(t-M^2)}$ and the Type-I open string amplitude} 
\label{fig_seftplot}
\end{figure}

We can test the bound against two candidate amplitudes. The first is given as 
\begin{equation}
M(s,t)=-\frac{1}{(s{-}M^2)(t{-}M^2)}\, .
\end{equation}
Note that the above will have an infinite number of higher spin particles with the same mass $M$, and thus does not make sense physically. However it can be shown that this amplitude satisfies our requirements, in particular the residue in the $s$-channel has a positive expansion, see\cite{Caron-Huot:2016icg}:
\begin{equation}
-\frac{1}{(s-M^2)(t-M^2)}=\frac{1}{s-M^2}\sum_{\ell\in\mathbb{Z}^+}p_{M,\ell}P_\ell\Big(1+\frac{2t}{M^2}\Big)
\end{equation}
where
\begin{equation}
p_{M,\ell}=\frac{2\ell+1}{M^2}\int_{-1}^0\mathrm{d}x\frac{P_\ell(1+2x)}{1-x}>0
\end{equation}
Thus $\langle p_{k,\ell}\rangle$ of this model is,
\begin{equation}
\langle p_{k,\ell}\rangle=\frac{2\ell+1}{M^{2(k+1)}}\int_{-1}^0\mathrm{d}x\frac{P_\ell(1+2x)}{1-x}\,.
\end{equation}
The second is the type-1 open string amplitude
\eq
M(s,t)=\frac{\Gamma[{-}s]\Gamma[{-}t]}{\Gamma[1{-}s{-}t]}\,,
\eqe
where the residue is given as
\eq
\text{Res}_n(t)=\frac{\prod_{i=1}^{n-1}(t+i)}{(n!)}\,.
\eqe
Since the higher level contributions are suppressed, for each $\langle p_{k,\ell}\rangle$ we only sum over contributions from the first 50 levels after the corresponding spin appears in the leading trajectory. In Figure \ref{fig_seftplot} we show that both $s$-matrix will satisfy the bounds in eq.(\ref{eqn:seftk5}).

\noindent \textbf{The full EFThedron}

Based on the analysis of previous section \ref{sec:spin spectrum}, we know that for identical scalars coupled to massive spin states, all the even spin particles must exist, for $\ell\le 28$. Thus it will be interesting to consider the ratio of $\langle p_{k,\ell}\rangle$ for any higher spin state with respect to spin 2.  We first look at $k=4,5,6$ case where one only has one null constraint. From \eqref{eqn:sunull} we get,
\begin{align}
&n_4=\sum_{\ell\in\text{even}}\langle p_{4,\ell}\rangle\ell(\ell+1)(\ell^2+\ell-8)=0\notag\\
&n_5=\sum_{\ell\in\text{even}}\langle p_{5,\ell}\rangle\ell (\ell+1) (\ell (\ell+1) (2\ell (\ell+1)-43)+150)=0\notag\\
&n_6=\sum_{\ell\in\text{even}}\langle p_{6,\ell}\rangle(\ell-3) \ell (\ell+1) (\ell+4) \left(\ell (\ell+1) \left(\ell^2+\ell-32\right)+204\right)=0
\end{align}
The sign pattern of these three null constraints is,
\begin{center}
\begin{tabular}{ |c|c|c|c|c|c|c|c|c| } 
 \hline
$\ell$	&	0	&	2	&	4	&	6	&	8	&\ldots \\ 
 \hline
$\tilde{\omega}_4(\ell)$&	0	&	-	&	+&	   +  	&	+ & \ldots	\\ 
 \hline
 $\tilde{\omega}_5(\ell)$&	0	&	-	&	+&	   +  	&	+ & \ldots	\\ 
 \hline
$\tilde{\omega}_6(\ell)$&	0	&	-	&	- &	   +   	&	+ & \ldots	\\ 
 \hline
\end{tabular}
\end{center}
Thus for $k=4,5$, we can use derive a valid upper bounds on $\langle p_{k,\ell}\rangle/\langle p_{k,2}\rangle$,
\begin{align}
&\frac{\langle p_{4,\ell}\rangle}{\langle p_{4,2}\rangle}\leq\frac{12}{\ell (\ell+1) \left(\ell^2+\ell-8\right)},~(\ell\geq4)\notag\\
&\frac{\langle p_{5,\ell}\rangle}{\langle p_{5,2}\rangle}\leq\frac{216}{\ell (\ell+1) (\ell (\ell+1) (2\ell (\ell+1)-43)+150)},~(\ell\geq4)\,.
\end{align}
For $k=6$, since there are two spins with negative value, we can only obtain bounds of the form similar to the $s$-EFThedron.

Starting from $k\geq7$, one will get more than one null constraint. In order to get the strongest upper bound for $\langle p_{k,\ell_a}\rangle/\langle p_{k,2}\rangle$, we require the special functional to satisfy the following conditions,
\begin{align}\label{eqn:funcon}
&\text{For}~\ell\geq4,~l\in\text{even},~\tilde{\omega}_k(l)\geq0\notag\\
&\text{Minimize}~-\frac{\tilde{\omega}_k(2)}{\tilde{\omega}_k(l_a)}
\end{align}
We call this kind of the special functional as optimal functional. At $k=7,9$, one will have two null constraints. The optimal functionals take the form,
\begin{equation}
\tilde{\omega}_{7}(\ell)=\det[\vec{G}^{(7)}_4,\vec{G}^{(7)}_{\ell}],\quad\tilde{\omega}_{9}(\ell)=\det[\vec{G}^{(9)}_4,\vec{G}^{(9)}_{\ell}]
\end{equation}
where vector $\vec{G}^{(k)}_\ell$ made by the independent null constraints at the level of $k$. The sign patterns are,
\begin{center}
\begin{tabular}{ |c|c|c|c|c|c|c|c|c| } 
 \hline
$\ell$	&	0	&	2	&	4	&	6	&	8	&\ldots \\ 
 \hline
$\tilde{\omega}_7(\ell)$&	0	&	-	&	0&	   +  	&	+ & \ldots	\\ 
 \hline
 $\tilde{\omega}_9(\ell)$&	0	&	-	&	0&	   +  	&	+ & \ldots	\\ 
 \hline
\end{tabular}
\end{center}
The upper bounds are,
\begin{small}
\begin{align}\label{eqn:suplbound}
&\frac{\langle p_{7,\ell}\rangle}{\langle p_{7,2}\rangle}\leq\frac{604800}{(\ell-4) \ell (\ell+1) (\ell+5) (\ell (\ell+1) (7\ell (\ell+1) (2\ell
   (\ell+1)-115)+11112)-33516)},~(\ell\geq6)\notag\\
&\frac{\langle p_{9,\ell}\rangle}{\langle p_{9,2}\rangle}\leq\frac{152409600}{(\ell-4) \ell (\ell+1) (\ell+5) (\ell (\ell+1) (\ell (\ell+1) (\ell (\ell+1)}\notag\\
&\times\frac{1}{(\ell (\ell+1)
   (2 \ell (\ell+1)-331)+19014)-448776)+3862728)-8899632)},\quad(\ell\geq6)
\end{align}
\end{small}
Starting from $k>11$, one cannot find any functional that gives bounds on $\langle p_{k,\ell}\rangle/\langle p_{k,2}\rangle$. But one can mange to get functional that give upper bound on $\langle p_{k,\ell}\rangle/\langle p_{k,\ell*\geq4}\rangle$.

\begin{figure}
\centering
\includegraphics[scale=0.45]{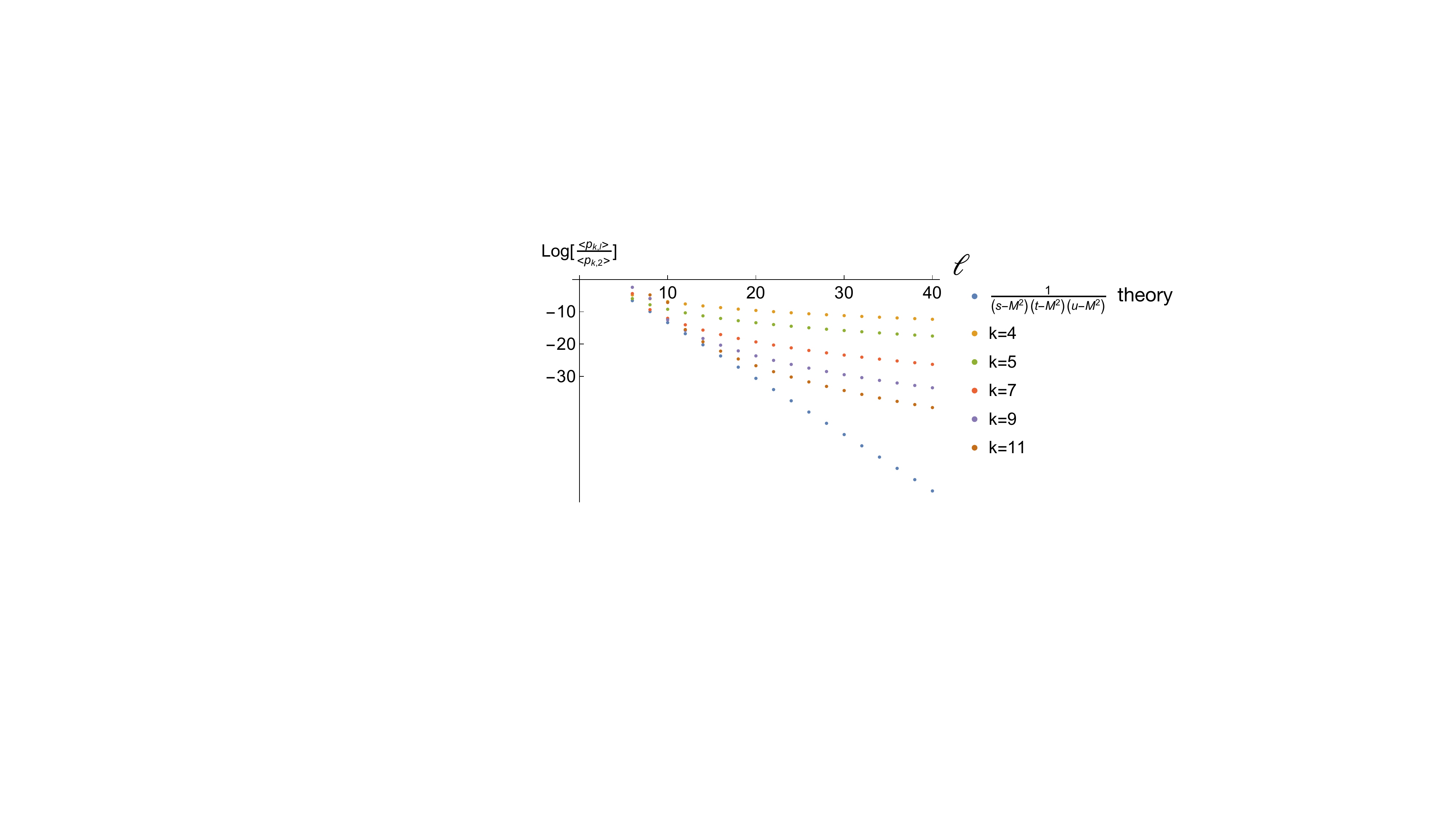}
\caption{Here we compare the different $k$ level upper bounds of $\frac{\langle p_{k,\ell}\rangle}{\langle p_{k,2}\rangle}$ in the $su$-EFThedron with the $S$-matrix $\frac{1}{(s-M^2)(t-M^2)(u-M^2)}$}
\label{fig_sueftplot}
\end{figure}

We can also test the bound using the following two $S$-matrices. The first has the simple form
\begin{equation}\label{stu}
M(s,t,u)=\frac{1}{(s-M^2)(t-M^2)(u-M^2)}\,.
\end{equation}
The $\langle p_{k,\ell}\rangle$ in this case is
\begin{equation}
\langle p_{k,\ell}\rangle=-\frac{2(2\ell+1)}{M^{2k-2}}\int_{-1}^1\mathrm{d}y\frac{1}{y^2-9}P_\ell(y)
\end{equation}
In Figure \ref{fig_sueftplot}, one can see that the 
average spectral function of this $S$-matrix is decaying exponentially as $\ell\to\infty$, while the bounds eq.\eqref{eqn:suplbound} are just polynomially suppressed.

\begin{figure}[h]
	\centering          
	\includegraphics[scale=0.45]{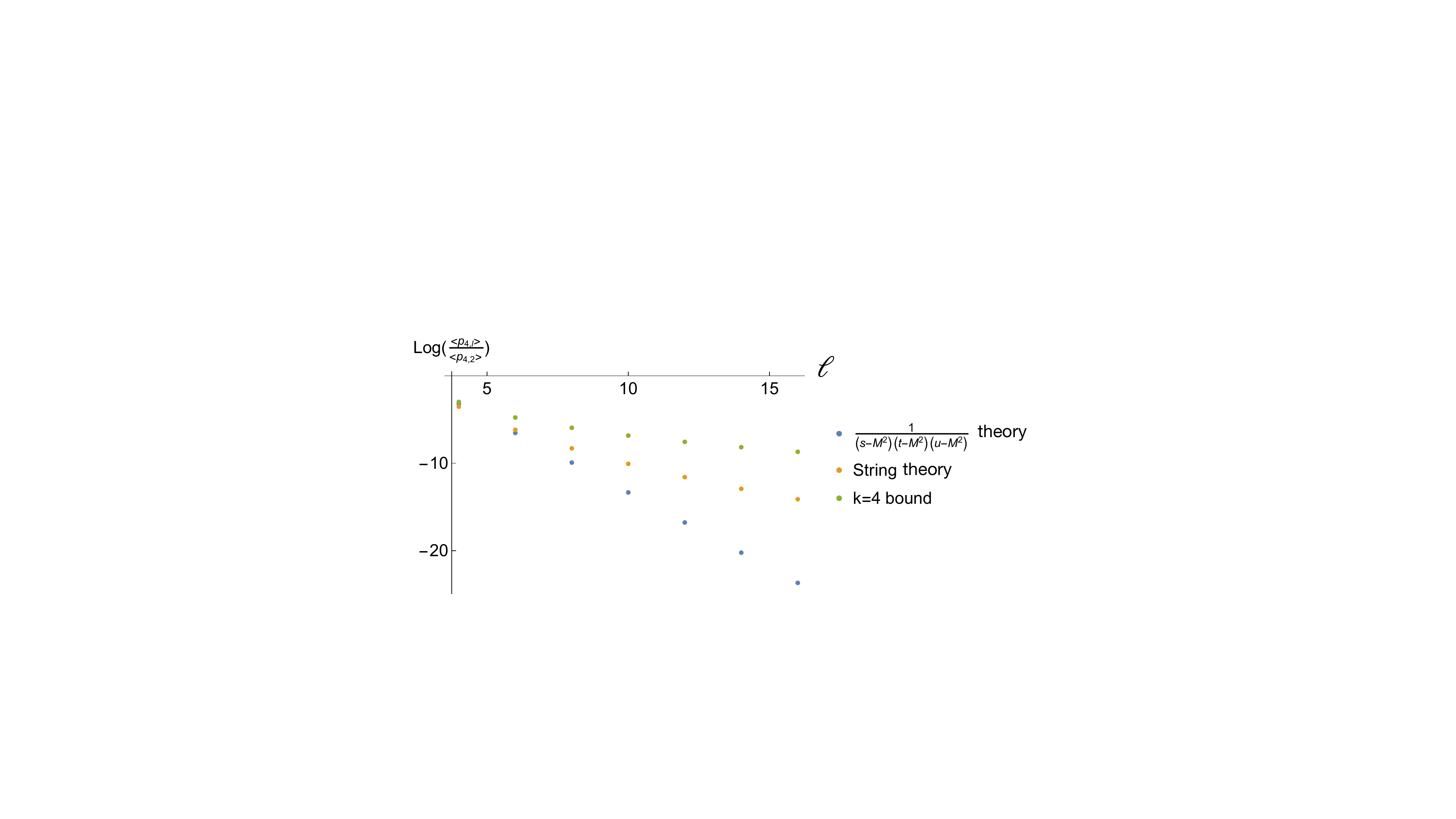}   
	\includegraphics[scale=0.45]{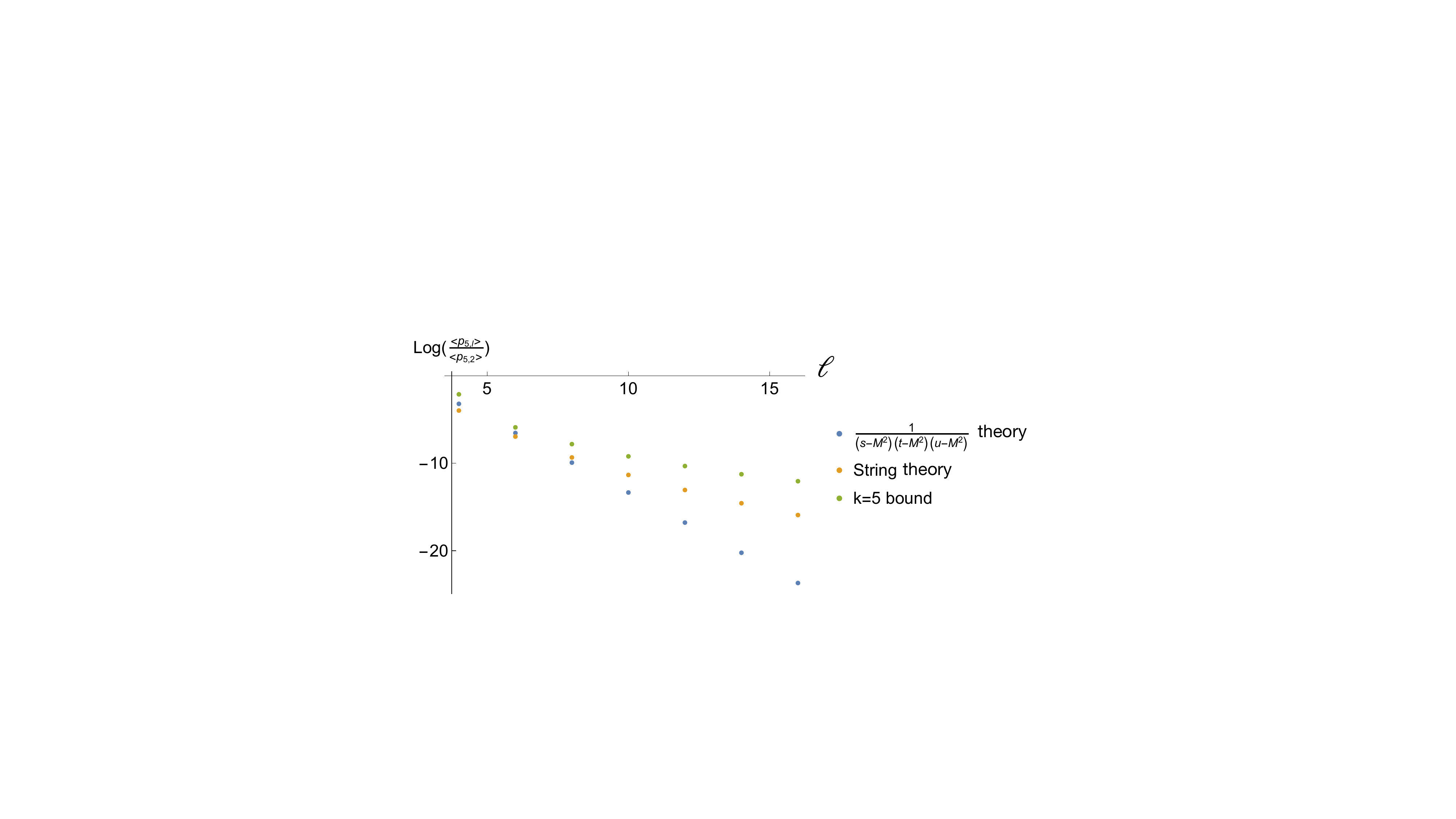}   	
\caption{Here we compare the $k=4,5$ upper bounds of $\frac{\langle p_{k,\ell}\rangle}{\langle p_{k,2}\rangle}$ with the $S$-matrix $\frac{1}{(s-M^2)(t-M^2)(u-M^2)}$ and the closed string theory}  
\label{fig:closedstring} 
\end{figure}

The second one is the closed-string amplitude,
\begin{equation}
M_{\text{string}}(s,t,u)=\frac{\Gamma({-}s)\Gamma({-}t)\Gamma({-}u)}{\Gamma(1{+}s)\Gamma(1{+}t)\Gamma(1{+}u)}\,,
\end{equation}
whose residues are just the square of the open string residues:
\begin{equation}
\text{Res}_n(t)=\frac{\prod_{i=1}^{n-1}(t+i)^2}{(n!)^2}\,.
\end{equation}
The result is displayed in Figure \ref{fig:closedstring}, where we see that it lies between our bound and the result for eq.(\ref{stu}).

\subsection{Implications for mass spectrum}
Previously, we have considered the null constraints each $k$ at a time. This imposes bounds on the spin content of the spectrum. Here instead we consider various linear combinations of null constraints from different $k$, which will result in constraints on the mass content.

\begin{figure}
\centering
\includegraphics[scale=0.65]{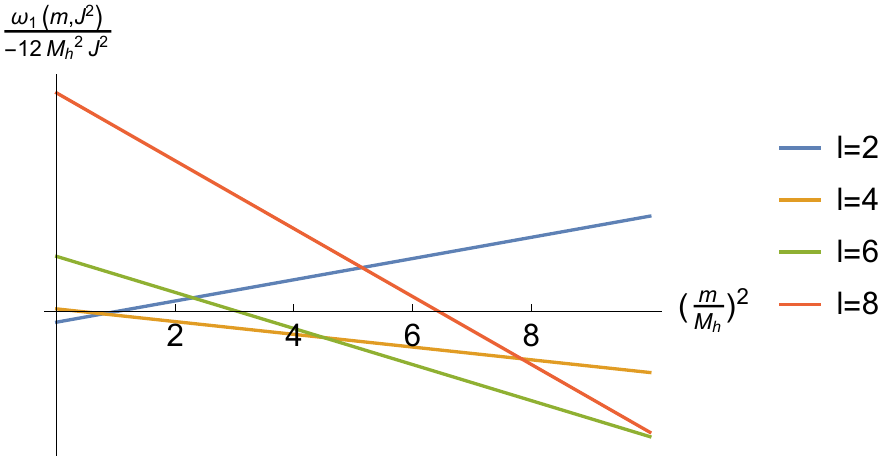}
\caption{Here we present the functional $\omega_1(m,J^2)$ for different spin. The position of the largest zero eq.\eqref{eqn:massgap1} will increase as $J^2$ increase.}
\label{fig_massgap}
\end{figure}

Let us begin with the $k=4,5$ null constraint and combine it into a vector equation,
\begin{equation}\label{eqn:massgapsumrule}
\begin{pmatrix}
n_4\\
n_5
\end{pmatrix}
=
\sum_{i}p_{i}
\begin{pmatrix}
\frac{J_i^2(J_i^2-8)}{m_i^{10}}\\
\frac{J_i^2(150+J_i^2(2J_i^2-43))}{m_i^{12}}
\end{pmatrix}
=\sum_{i,x}\frac{p_{i}}{m_i^{12}}
\begin{pmatrix}
J_i^2(J_i^2-8)m_i^2\\
J_i^2(150+J_i^2(2J_i^2-43))
\end{pmatrix}
=0
\end{equation}
Now we assume that at $\ell=2$, the mass are bounded from above, and the mass of heaviest spin-$2$ particle is $M_h$. Consider the following functional,
\begin{equation}\label{eqn:massgapfun1}
\omega_1(m,J^2)=\det[\vec{G}_{M_h,6},\vec{G}
_{m,J^2}]=-12J^2M^2\Big[150-43J^2+2J^4-18(J^2-8)\Big(\frac{m}{M_2}\Big)^2\Big]
\end{equation}
where 
\begin{equation}
\vec{G}_{m_i,J^2_i}=
\begin{pmatrix}
J_i^2(J_i^2-8)m_i^2\\
J_i^2(150+J_i^2(2J_i^2-43))
\end{pmatrix}
\end{equation}
By construction the functional will have a zero exactly at the largest massive spin 2 state. The plot of eq.\eqref{eqn:massgapfun1} for each spin is at Figure \ref{fig_massgap}. For $\ell>2$ the function has a negative slope and only for spin $2$ does the slope becomes positive. Since the zero occurs at the largest mass amongst the spin 2 states,  all $\ell=2$ particles  have positive contribution. Thus in order for the sum rule to be true, \emph{there must exist a higher spin particle with mass} 
\begin{equation}\label{eqn:massgap1}
m^2_\ell\leq\frac{2J^4- 43 J^2+150}{18 J^2-144 }M_h^2\,.
\end{equation}

This bound becomes stronger by adding $k=6$ constraint. The functional in this case is,
\begin{eqnarray}\label{eqn:massgapfun2}
\omega_2(m,J^2)&=&\det[\vec{G}_{M_h,6},\vec{G}_{\sqrt{5}\frac{M}{2},42},\vec{G}
_{m,J^2}]\nonumber\\
&\sim&\Big[(12{-}J^2) ((J^2{-}32)J^2{+}204){+}8 (J^2 (2J^2{-}43){+}150) \Big(\frac{m}{M_h}\Big)^2\Big]
\end{eqnarray}
where the $\vec{G}_{m,J^2}$ now is determined by combining three null constraints,
\begin{equation}
\vec{G}_{m,J^2}=
\begin{pmatrix}
m^4J^2(J^2-8)\\
m^2J^2(150+J^2(2J^2-43))\\
J^2(J^2-12)(J^2(J^2-32)+204)
\end{pmatrix}
\end{equation}
Using the same method, the mass gap now can be improved as,
\begin{equation}
m^2_\ell\leq M_h^2\frac{( J^2-12) ( J^4  - 32 J^2 +204)}{8 (150 - 43 J^2 + 2 J^4)}
\end{equation}

Note that the dominant contribution to $\langle p_{k,\ell}\rangle$ is the lightest state of a given spin. Thus an upper bound on $\langle p_{k,\ell}\rangle$ is a constraint on this lightest state, which cannot become arbitrary light without having its coupling $p_{\ell_i}$ being heavily suppressed. Here the bound is in the opposite direction, where it requires some state in the spectrum must be sufficiently ``light". Of course, if the masses of the spin 2 states are unbounded, the condition is less useful. However, the simplicity of these results suggests the complete geometry in $(p,\ell,m)$ space is highly non-trivial.

\subsection{Constraints on EM and gravitational EFTs}
In this section we briefly consider some of the above analyses when the external scalars are replaced by massless particles with spin, leading to constraints for  UV completions of electromagnetic or gravitational EFTs. For external helicity states, dispersion relations for the couplings of an amplitude $M({+}h,{+}h,{-}h,{-}h)$ are given as~\cite{Arkani-Hamed:2020blm, Bern:2021ppb}:
\eqa
 [12]^{2h}\langle 34\rangle^{2h}(\sum_{k,q}g_{k,q}s^{k{-}q}t^q)={-}[12]^{2h}\langle 34\rangle^{2h}\left(\sum_{\ell_a\geq0} p_{i} \frac{d_{0,0}^{\ell_i{=}even}(\theta)}{s{-}m^2_i}{+}\sum_{\ell_j\geq2h} p_j \frac{\tilde{d}_{2h,2h}^{\ell_j}(\theta)}{{-}t{-}s{-}m^2_j}\right)\nonumber\\
\eqae
where $d^{\ell}_{a,b}(\theta)$ is the Wigner-$d$ matrix, while $\tilde{d}_{2h,2h}^{\ell}(\theta)\equiv d_{2h,2h}^{\ell}(\theta)/(\cos \theta/2)^{4h}$, and the $s$-channel residue only allows for even spin. Once again the equality is understood as a Taylor expansion on both sides. Since the spinor bracket prefactor behaves as $s^{2h}$, the dispersive representation is valid for any $k,q$. The Wigner $d$-matrices can be expanded in small $t$ by $\cos(\theta)=1+2 t$ and expand in $t$. It turns out that $d_{0,0}^{\ell=even}(\theta)$ is just our usual Legendre polynomial so its expansion written in $J^2=\ell(\ell+1)$ is as before. For $\tilde{d}_{2,2}^{\ell}(\theta)$, we have
\eq
\tilde{d}_{2h,2h}^{\ell}(\theta)\bigg|_{\theta=\cos^{-1}(1+2t)}=\sum_{q}v_{\ell,q}t^q\,,
\eqe
with 
\eq
v_{\ell,q}=\frac{\prod_{a=1}^q(J^2-(a+2h)(a+2h-1))}{(q!)^2}\,.
\eqe

Note that our helicity configuration is invariant under $1\leftrightarrow 2$ exchange. Since our dispersive representation for the couplings are valid for any $k,q$, the constraint that this symmetry imposes on the spectrum can be obtained by Taylor expanding, 
\begin{align}
M(s,t,u){-}M(s,u,t)= \left(\sum_{\ell_i\geq0} p_{i} \frac{d_{0,0}^{\ell_i{=}even}(\theta)}{s{-}m^2_i}{-}\sum_{\ell_j\geq2h} p_j \frac{\tilde{d}_{2h,2h}^{\ell_j}(\theta)}{t{+}s{+}m^2_j}\right){-}(t\leftrightarrow u)=0\,.
\end{align}
The constraint is then collecting the coefficients of each $s^at^b$, which must be zero. The null constraint is at level $k$ is,
\begin{equation}
\sum_{\ell\in\text{even}}\langle p^{s}_{\ell,k}\rangle\omega_k^s(\ell)+\sum_{\ell\geq2h}\langle p^{u}_{\ell,k}\rangle\omega_k^u(\ell)=0
\end{equation}
where $\langle p^{s}_{\ell,k}\rangle$ and $\langle p^{u}_{\ell,k}\rangle$ are the average spectral function in $s$ and $u$-channels. The remaining steps are identical with that of the scalar EFThedron. In the following we simply do the analysis for $k=1,2$ when there is only one null constraint. 
\begin{align}
&\sum_{l\in\text{even}}\langle p_{l,1}^s\rangle\ell(\ell+1)+\sum_{l\geq 2h}\langle p_{\ell,1}^u\rangle (-4 h^2-2 h+\ell^2+\ell-1)=0\notag\\
&\sum_{l\in\text{even}}\langle p_{l,2}^s\rangle\ell(\ell+1)(\ell+\ell^2-6)+\sum_{l\geq 2h}\langle p_{\ell,2}^u\rangle 4-(2 h-\ell) (2 h-\ell+1) (2 h+\ell+1) (2 h+\ell+2)=0
\end{align}
Notice that $\tilde{\omega}^s(\ell)$ are non-negative for $\ell$ given by even integers. Thus we just have to look at the sign pattern for $\tilde{\omega}^u(\ell)$,
\begin{center}
\begin{tabular}{ |c|c|c|c|c|c|c|c|c| } 
 \hline
$\ell$	&	2	&	3	&	4	&	5	&	6 & 7	&\ldots \\ 
 \hline
$h=1,\tilde{\omega}^u_1(\ell)$&	-	&	+	&	+ &	   +  	&+&	+ & \ldots	\\ 
 \hline
$h=2,\tilde{\omega}^u_1(\ell)$&		&		&	-&	   + &	+ &+& \ldots	\\ 
 \hline
$h=1,\tilde{\omega}^u_2(\ell)$&	-	&	-	&	+ &	   +  	&+&	+ & \ldots	\\ 
\hline
$h=2,\tilde{\omega}^u_2(\ell)$&		&		&	-&	   - &	+ &+& \ldots	\\ 
 \hline

\end{tabular}
\end{center}
Thus when $k=1$, we know that there must be spin 2 and spin 4 particles in the $u$-channels in the EM and gravitational EFTs respectively. We can then derive upper bounds for average spectral functions in both $s$ and $u$-channels,
\begin{align}
&h=1:\quad\frac{\langle p_{\ell,1}^s\rangle}{\langle p_{2,1}^u\rangle}\leq\frac{1}{\ell(\ell+1)}~(\ell\geq2),\quad\frac{\langle p_{\ell,1}^u\rangle}{\langle p_{2,1}^u\rangle}\leq\frac{1}{\ell^2+\ell-7}~(\ell\geq3)\notag\\
&h=2:\quad\frac{\langle p_{\ell,1}^s\rangle}{\langle p_{4,1}^u\rangle}\leq\frac{1}{\ell(\ell+1)}~(\ell\geq2),\quad\frac{\langle p_{\ell,1}^u\rangle}{\langle p_{4,1}^u\rangle}\leq\frac{1}{\ell^2+\ell-21}~(\ell\geq5)
\end{align}
When $k=2$, from sign pattern table we know that there must be spin 2 or 3 particles and EM EFTs and spin 4 or 5 particles gravitational EFTs. Thus in this case we cannot get an upper bound for $\langle p_{2,\ell}^{s,u}\rangle$ with respect to only a single average spectral function. The results are,
\begin{align}
&h=1:\quad\langle p_{l,2}^s\rangle\leq4\frac{\langle p_{2,2}^u\rangle+\langle p_{3,2}^u\rangle}{\ell(\ell+1)(\ell^2+\ell-6)}~(\ell\geq4),\quad\langle p_{l,2}^u\rangle\leq4\frac{\langle p_{2,2}^u\rangle+\langle p_{3,2}^u\rangle}{68+\ell(\ell+1)(\ell^2+\ell-18)}~(\ell\geq4)\notag\\
&h=2:\quad\langle p_{l,2}^s\rangle\leq4\frac{\langle p_{4,2}^u\rangle+\langle p_{5,2}^u\rangle}{\ell(\ell+1)(\ell^2+\ell-6)}~(\ell\geq4),\quad\langle p_{l,2}^u\rangle\leq4\frac{\langle p_{4,2}^u\rangle+\langle p_{5,2}^u\rangle}{596+\ell(\ell+1)(\ell^2+\ell-50)}~(\ell\geq6)
\end{align}




\section{Conclusion and outlook} \label{conc}
In this paper, we have explored in detail the analytic description of the boundaries for the space of consistent EFT. This space is defined through the dispersive representation of the Wilson coefficients. We have demonstrated that at the core of the EFThedron, first introduced in~\cite{Arkani-Hamed:2020blm}, is the geometry given by the convex hull of product moment curves. In particular, we have the product of a continuous bounded curve, associated with the mass spectrum of the UV completion, and a discrete unbounded moment associated with the spin content of the spectrum. The identification of this geometry allows us to state the boundary of the EFThedron in the limit of infinite dimensions, which corresponds to coefficients of higher dimension operators to all orders in derivatives.

In a generic setup, we only have access to a finite number of Wilson coefficients. In such case the boundaries in the infinite dimension limit are no longer complete. Nevertheless, even for the ``truncated" geometry we present a finite collection of constraints that is very close to fully describing the space, and converges to the exact solution in the infinite limit. The space of consistent EFTs is then carved out by considering the intersection of the symmetry plane, defined either by cyclic or permutation invariance of the amplitude, and the EFThedron. We have compared the space carved out from the geometry with that from SDP, with the results in perfect agreement in the $s-u$ case, where conditions are slightly simpler to solve. For the $s$-EFThedron, we expect including higher order constraints will similarly reduce the region to match the functional method.

In general the intersection yields a finite region. The reason can be attributed to the fact that for any symmetry plane, there are only finite number of spins on one side of the plane, while the majority is on the other side. This implies non-trivial constraint on the UV spectrum. In particular we demonstrate that for permutation invariant amplitudes, the UV completion \textit{must} contain all even spins $\ell\leq 28$. We also derived upper bounds on the ratios of averaged spinning spectral function, which demonstrates the phenomenon of large spin suppression. Finally, we have shown that at least one high spin particle satisfy an upper bound $m^2_\ell\le M^2 B(\ell)$, where $M^2$ is the mass of the heaviest spin 2. All of these results have their generalization to the case when the external states carry spins, i.e. photons and gravitons. Much like the EFThedron, the UV parameters $\{p,\ell,m\}$ form a highly non-trivial geometry (though no longer convex), induced by the null conditions. We leave a more systematic study of this problem to the future.

In all of the previous discussion, we have only required that $p_i$ is positive. However,  the spectral function is further bounded by unitarity to lie between $[0,2]$. Indeed by exploiting this upper limit, the authors of \cite{Caron-Huot:2020cmc} were able to obtain upper bounds for $g_{2,0}$ in the case of identical scalars. It is an open problem how to implement such constraint in the geometry, as well as in the numeric SDP setup. 

Generically, one has more than one massless degree of freedom in an EFT. It will be interesting to explore the constraint of positivity for the scattering of arbitrary mixed states, and in particular, their implications on the UV spectrum. Some analysis along these lines have recently been pursued in~\cite{Arkani-Hamed:2020blm, Zhang:2020jyn, Bern:2021ppb, Li:2021cjv}. 

Now that the geometry of the EFThedron is settled, and new boundaries have been obtained by using the moment matrix, it would be interesting if the analysis performed in \cite{Huang:2020nqy} can be improved by obtaining a stronger convergence to the open string solution. As in this paper we have also obtained the complete boundary structure for the $su$-EFThedron, similar analyses may now be carried out for closed strings (see also \cite{Green:2019tpt}). One may also consider null constraints associated to BCJ relations, which can be used to bootstrap EFTs~\cite{Carrasco:2019yyn}.

We have seen that making contact with the mathematical moment problem was a crucial step in better understanding the space of physical EFTs. However, many other possibly fascinating connections to related results remain unexplored. The operations of slicing and projecting the EFThedron are central, but they spoil the simple boundary structure of Hankels or cyclic polytopes. Describing the resulting shapes of slices and shadows of polytopes has also been previously investigated \cite{10.1145/237218.237228}.  As we have mentioned previously, several aspects of the truncated moment problem are still open problems. For instance, a much more efficient method to determine the boundary of the truncated discrete moment problem was recently proposed in \cite{INFUSINO2017443}. While not directly applicable to our case, as the discrete moments are valued in $\{0,1,2,\ldots\}$, it suggests new closed form solutions may exist for various problems.

The generalization to higher dimensions is straightforward for external scalars. As mentioned in the introduction, changing from Legendre to Gegenbauer polynomials is reflected merely in a different GL transformation quoted in eq.(\ref{ExplicitGL}). For external massless spins, the situation is more complicated, as one can now exchange various mixed representation of SO($D{-}1$), and the three-point couplings for such exchange are no-longer unique, albeit constrained from gauge invariance. Thus one in general have an $n\times n$ matrix of polynomials, where $n$ is the number of three-point couplings, and unitarity is reflected in the coefficient matrix being positive semi-definite. The optimal geometric description of such constraints remains to be explored.

Finally, the same convex hull of product moment curve also underlies the torus partition function of 2D CFTs~\cite{ModularHedron}, where the Taylor coefficients are associated with the expansion of the partition function around the self-dual fix point under modular invariance. The symmetry subplane is then derived from modular invariance. This suggests that there is a formal ``duality" between the space of consistent EFT and that of 2D torus partition functions. It will be interesting to explore this connection in more detail. Similarly it would be interesting to investigate the connections between the EFTs as those coming from the flat space limit of AdS/CFT, where the AdS EFT is constrained by the boundary CFT. For recent results see~\cite{Kundu:2021qpi}.

\acknowledgments 
We would like to thank Ting-Chun Lin for useful discussions on the multi-variate moment problem, and Nima Arkani-Hamed for suggestions on the spin-suppression analysis. L-y Chiang, Y-t Huang, and H-c Weng are supported by Taiwan Ministry of Science and Technology Grant No. 109-2112-M-002 -020 -MY3.   L Rodina is supported by Taiwan Ministry of Science and Technology Grant No. 109-2811-M-002-523.

\appendix 

\section{Using Semi-definite programming to carve out the space of EFT coefficients}
In \cite{Caron-Huot:2020cmc}, the authors shows that one can use semi-definite programming to help carve out the space of EFT coefficients. Here we briefly review their method for self-containedness.

\noindent \textbf{One parameter space (Bounds on $\tilde{g}_{k,q}$)}

If a EFT coefficient $g_{k,q}$ has a dispersive representation, we know that generally it has the form,
\begin{equation}
g_{k,q}=\sum_i p_{m_i,\ell_i}\frac{F_{k,q}(\ell)}{m_i^{2(k+1)}}
\end{equation}
where $f_{k,q}(\ell)$ is polynomial of $\ell$. Also because the EFThedron is constrained to a sub-plane, which will generate infinite number of null constraints which has the form,
\begin{equation}
n_k=\sum_i p_{m_i,\ell_i}\frac{B_k(\ell)}{m_i^{2(k+1)}}=0
\end{equation}
where $k$ labels the level of null constraints. With these two ingredients in hand we can transform it to semi-definite programming problem. Let us define the vector $\vec{E}_{m,\ell}$ as,
\begin{equation}
\vec{E}_{m,\ell}=
\begin{pmatrix}
\frac{1}{m_i^{6}}\\
\frac{M^{2(k-2)}F_{k,q}(\ell)}{m_i^{2(k+1)}}\\
\frac{B_{k_1}(\ell)}{m_i^{2(k_1+1)}}\\
\frac{B_{k_2}(\ell)}{m_i^{2(k_2+1)}}\\
\vdots
\end{pmatrix}
\end{equation}
Thus we can write down a vector equation,
\begin{equation}
\sum_{m_i,\ell_i}p_{m_i,\ell_i}\vec{E}_{m_i,\ell_i}=
\begin{pmatrix}
1\\
\tilde{g}_{k,q}\\
0\\
0\\
\vdots
\end{pmatrix}
\end{equation}
Then we redefine the $\tilde{p}_{m_i,\ell_i}$ as,
\begin{equation}
\tilde{p}_{m_i,\ell_i}=\frac{p_{m_i,\ell_i}}{m_i^{2(\tilde{k}+1)}},\quad\text{where}~\tilde{k}=\text{max}(k,k_1,k_2,\dots)
\end{equation}
And make the substitution $m_i^2\to M^2(1+x_i)$. The vector equation now becomes,
\begin{equation}
\sum_{x_i>0,\ell_i}\tilde{p}_{x_i,\ell_i}\vec{E}_{x_i,\ell_i}=
\begin{pmatrix}
1\\
\tilde{g}_{k,q}\\
0\\
0\\
\vdots
\end{pmatrix}
\end{equation}
Notice that now each entry in the vector $\vec{E}_{x_i,\ell_i}$ is just polynomial in $x_i$, with $x_i>0$. The corresponding semi-definite programming problems are,
\begin{itemize}
\item Upper bound ($\tilde{g}_{k,q}\leq A$)
\begin{align}
&\text{Minimize}\quad A=\vec{\alpha}\cdot(1,0,0,\cdots)\notag\\
&\text{Such that}\quad\text{For}~x\geq0,~\vec{\alpha}\cdot\vec{E}_{x,\ell}\geq0,\quad\vec{\alpha}\cdot(0,-1,0,0,\dots)=1
\end{align}
\item Lower bound ($B\leq\tilde{g}_{k,q}$)
\begin{align}
&\text{Maximize}\quad B=\vec{\alpha}\cdot(-1,0,0,\cdots)\notag\\
&\text{Such that}\quad\text{For}~x\geq0,~\vec{\alpha}\cdot\vec{E}_{x,\ell}\geq0,\quad\vec{\alpha}\cdot(0,1,0,0,\dots)=1
\end{align}
\end{itemize}

\noindent \textbf{Two parameter space ($\tilde{g}_{k_1,q_1}$, $\tilde{g}_{k_2,q_2}$)}

The strategy is that one can first use the method of the previous section to get the upper and lower bound of $\tilde{g}_{k_1,q_1}$
\begin{align}
A_1\leq\tilde{g}_{k_1,q_1}\leq A_2
\end{align}
Then at each value $\beta\in[A_1,A_2]$ we use semi-definite programming again to get the upper and lower bound of $\tilde{g}_{k_2,q_2}$,
\begin{equation}
\text{For each}~\beta,~B_1(\beta)\leq\tilde{g}_{k_2,q_2}\leq B_2(\beta)
\end{equation}
Then we can carve out the 2D space of ($\tilde{g}_{k_1,q_1}$, $\tilde{g}_{k_2,q_2}$). Below we show the optimization problem at some $\beta$. The vector $\vec{E}_{x_i,\ell_i}$ in this problem is,
\begin{equation}
\sum_{x_i\geq0,\ell_i}\tilde{p}_{x_i,\ell_i}\vec{E}_{x_i,\ell}=
\begin{pmatrix}
1\\
\tilde{g}_{k_1,q_1}\\
\tilde{g}_{k_2,q_2}\\
0\\
0\\
\vdots
\end{pmatrix}
\end{equation}
\begin{itemize}
\item  Upper bound
\begin{align}
&\text{Minimize}\quad B_2(\beta)=\vec{\alpha}\cdot(1,\beta,0,0,\cdots)\notag\\
&\text{Such that}\quad\text{For}~x\geq0,~\vec{\alpha}\cdot\vec{E}_{x,\ell}\geq0,\quad\vec{\alpha}\cdot(0,0,-1,0,\dots)=1
\end{align}
\item  Lower bound
\begin{align}
&\text{Maximize}\quad B_1(\beta)=\vec{\alpha}\cdot(-1,\beta,0,0,\cdots)\notag\\
&\text{Such that}\quad\text{For}~x\geq0,~\vec{\alpha}\cdot\vec{E}_{x,\ell}\geq0,\quad\vec{\alpha}\cdot(0,1,0,0,\dots)=1
\end{align}
\end{itemize}
This algorithm can be easily generalized to carving out higher dimensional parameter space. And for SDP solver we are using SDPB \cite{Simmons-Duffin:2015qma, Landry:2019qug} to perform our computation.

\bibliographystyle{JHEP}
\bibliography{refs.bib}

\end{document}